\documentclass[%
        12pt,
        notitlepage, 
        tightenlines,
        eqsecnum,
        floats,
        floatfix,
        nofootinbib,
        amsmath,amssymb,
        aps,prd,
        superscriptaddress,
        longbibliography]{revtex4-2}

\usepackage[utf8]{inputenc}
\usepackage{microtype}

\usepackage[normalem]{ulem}

\usepackage{mathtools}
\usepackage{graphicx}
\usepackage[caption=false]{subfig}
\usepackage{float}
\usepackage{amssymb}
\usepackage{xcolor}

\usepackage[hidelinks]{hyperref}

\def\dd{{\rm d}}
\def\f{\frac}

\def\lp{\ell_{\rm Pl}}

\definecolor{newgreen}{rgb}{0.0, 0.75, 0.0}

\begin{document}

\title{Gravitational collapse in effective loop quantum gravity: \\
beyond marginally bound configurations}

\author{Lorenzo Cipriani}
\email{lorenzo.cipriani@graduate.univaq.it}
\affiliation{Dipartimento di Scienze Fisiche e Chimiche, Università dell’Aquila, via Vetoio, I-67100, L’Aquila, Italy}
\affiliation{INFN, Laboratori Nazionali del Gran Sasso, I-67100 Assergi (AQ), Italy}
\author{Francesco Fazzini}
\email{francesco.fazzini@unb.ca}
\author{Edward Wilson-Ewing}
\email{edward.wilson-ewing@unb.ca}
\affiliation{Department of Mathematics and Statistics, University of New Brunswick, Fredericton, NB, Canada E3B 5A3}

\begin{abstract}

We study gravitational collapse in effective loop quantum gravity, focusing on non-marginally bound configurations in Lema\^itre-Tolman-Bondi spacetimes. In the homogeneous limit we recover the effective dynamics of loop quantum cosmology for Friedman cosmologies with spatial curvature. We study a particular family of configurations with a homogeneous interior and a sharp boundary where the dust energy density rapidly and continuously decreases to zero. For these configurations, the gravitational collapse continues to the Planck regime when a bounce occurs, at which point the dust ball starts to expand, and a shock wave forms in the gravitational field within the order of a Planck time after the bounce. The shock slowly moves outwards, eventually reaching the horizon which then disappears, at which time there is no longer a black hole. If the initial configuration is bound, the shock asymptotes to a maximal radius, whereas for unbound initial configurations the shock escapes to infinity. In all cases, the black hole lifetime is proportional to the square of the black hole mass, and additionally depends on how strongly bound the dust profile is; this last quantity also affects the vacuum region outside the dust profile which is not solely determined by the black hole mass and charge as in spherically symmetric general relativity. We also use numerics to study a wide range of other types of initial configurations, both bound and unbound, with qualitatively similar results.

\end{abstract}

\maketitle

\section{Introduction}

The physics of the gravitational collapse of a star, and the associated question of the ultimate fate of the star, is a topic of great interest in astrophysics. Ever since the groundbreaking work by Oppenheimer and Snyder \cite{Oppenheimer:1939ue}, significant efforts have been made to construct models and gain a deeper understanding of the underlying physics when a massive star undergoes gravitational collapse. It is now widely accepted that if the mass of the star is sufficiently large, it will eventually become a black hole. This prediction is based on Einstein's theory of gravity and applies in a regime where the theory is believed to be valid; however, the classical theory also predicts that such a collapse will inevitably lead to the formation of a singularity in the spacetime \cite{Penrose:1964wq}---singularities such as these are commonly thought to be pathologies of the classical theory that should be resolved by a complete theory of quantum gravity.

It is then natural to ask what the impact of quantum gravity may be on the process of gravitational collapse.  For a sufficiently large star, it seems reasonable to expect that quantum gravity effects will be negligible until well after the black hole has formed, and that they will only be large in regions where the spacetime curvature reaches the Planck scale.  Still, it is clearly of interest to develop models of gravitational collapse that include quantum gravity effects in order to answer this question more precisely, and in a quantitative manner.

Our goal in this paper is to study how holonomy corrections, motivated by loop quantum gravity (LQG), affect the dynamics of the gravitational collapse of dust, as captured in the Lema\^itre-Tolman-Bondi (LTB) spacetime that has been the subject of considerable study in LQG \cite{Bojowald:2008ja, Bojowald:2009ih, Tibrewala:2012xb, Liu:2014kra, Kelly:2020lec, Alonso-Bardaji:2021tvy, Husain:2021ojz, Husain:2022gwp,  Giesel:2023tsj, Giesel:2023hys, Fazzini:2023ova, Alonso-Bardaji:2023qgu}.

In addition to the studies of black holes in loop quantum gravity recently reviewed in \cite{Perez:2017cmj, Gambini:2022hxr, Ashtekar:2023cod}, a variety of approaches have been developed to study gravitational collapse in the context of loop quantum gravity. This includes models that have a finite number of degrees of freedom, like the Oppenheimer-Snyder model (corresponding to a star, with a homogeneous and isotropic interior, and vacuum exterior) \cite{Modesto:2006qh, Ziprick:2009nd, Tavakoli:2013rna, Rovelli:2014cta, Barrau:2014hda, Haggard:2014rza, Barcelo:2014npa, Barcelo:2014cla, Barrau:2014yka, Barrau:2015uca, Christodoulou:2016vny, Malafarina:2017csn, Olmedo:2017lvt, Bianchi:2018mml, Kelly:2020lec, BenAchour:2020bdt, BenAchour:2020mgu, BenAchour:2020gon, Munch:2020czs, Schmitz:2020vdr, Munch:2021oqn, Li:2021snn, Lewandowski:2022zce, Giesel:2022rxi, Han:2023wxg, Bobula:2023kbo, Fazzini:2023scu, Giesel:2023hys, Cafaro:2024} and thin-shell collapse models \cite{Hossenfelder:2009fc, Campiglia:2016fzp, Giesel:2021dug, Han:2022rsx} (see also \cite{Hayward:2005gi, Bambi:2013caa, Barcelo:2015uff, Ziprick:2016ogy, Schmitz:2019jct, BenAchour:2020bdt, Piechocki:2020bfo} for studies of these models in Wheeler-DeWitt and other approaches to quantum gravity), as well as richer models that allow for local degrees of freedom, typically by including dust \cite{Kelly:2020lec, Husain:2021ojz, Alonso-Bardaji:2021tvy, Husain:2022gwp,  Bojowald:2008ja, Bojowald:2009ih, Fazzini:2023ova, Giesel:2023tsj, Giesel:2023hys, Liu:2014kra, Alonso-Bardaji:2023qgu} or a scalar field \cite{Bojowald:2015zha, BenAchour:2016brs, Benitez:2020szx, Alonso-Bardaji:2020rxb, Gambini:2021uzf} in spherically symmetric gravity (see also \cite{Kreienbuehl:2010vc, Vaz:2011zz, Kiefer:2019csi, Hergott:2022hjm, Bonanno:2023rzk} for studies of such models in other approaches to quantum gravity).

Here, we will consider the model for the LTB spacetime, corresponding to dust minimally coupled to gravity in spherical symmetry. The effective dynamics for the LTB spacetimes that we use, including corrections from LQG, are derived by: (i) first at the classical level imposing spherical symmetry, using the dust field as a relational clock, and fixing the spatial diffeomorphism constraint by using the areal gauge, (ii) then holonomy corrections, due to the discreteness of quantum geometry predicted by LQG, are included in the resulting Hamiltonian from which the equations of motion can be derived; for details see \cite{Kelly:2020lec, Husain:2021ojz, Husain:2022gwp}. Further, it has recently been shown that it is not necessary to gauge-fix the diffeomorphism constraint before including holonomy corrections, rather this step can be done after including holonomy corrections \cite{Giesel:2023tsj, Giesel:2023hys}---in either case, the resulting LQG effective equations of motion for the LTB spacetime are the same. Note that the choices of the dust-time and areal gauges is equivalent to using generalized Painlev\'e-Gullstrand coordinates, a convenient choice that, already at the classical level, is known to be well-suited for a Hamiltonian analysis as well as for numerics \cite{Lasky:2006hq, Giesel:2009jp}.

In this paper, we will focus on configurations corresponding to gravitational collapse that are not marginally bound.  Collapsing configurations that are marginally bound are those for which all of the dust content, in the infinite past, was at infinity with vanishing velocity---in this sense, these are configurations for which the kinetic energy and gravitational potential energy of each shell composing the star sum to zero. In generalized Painlev\'e-Gullstrand coordinates, the marginally bound configurations have vanishing spatial curvature (though the spacetime curvature is non-vanishing). For the LQG effective equations of motion of interest here, the marginally bound case has already been studied in some detail \cite{Husain:2021ojz, Husain:2022gwp}, so as mentioned above here we will study configurations that are not marginally bound. Specifically, we will consider a range of initial configurations, including some where the dust is not sufficiently energetic to escape to infinity, and others where the dust can escape to infinity with a leftover velocity; these possibilities correspond, respectively, to positive or negative spatial curvature in generalized Painlev\'e-Gullstrand coordinates. (Note that for the Oppenheimer-Snyder model of collapse, these possibilities correspond to an interior that is, respectively, a closed or open Friedman universe.)

There are two intertwined main objectives in this work. The first is to check the robustness of the main results derived in the marginally bound case: do the main features persist beyond the particular family of marginally bound solutions? The second is to study the impact of spatial curvature on the results. This is especially important since only a few studies, such as \cite{BenAchour:2020mgu, Cafaro:2024}, have so far studied solutions to gravitational collapse in LQG beyond the spatially flat case.

Concerning the first main objective, numerical simulations starting from a wide range of initial configurations have shown that for the marginally bound case, LQG effects cause two main qualitative effects in gravitational collapse: a bounce occurs in the Planck regime, and a discontinuity in the gravitational field forms, typically shortly before or after the bounce (within a time of $\sim t_{\rm Pl}$) \cite{Husain:2021ojz, Husain:2022gwp}, although it is possible to choose initial conditions so the discontinuity forms well before the bounce. This discontinuity is a shock wave, which is a weak solution to the dynamics, and the shock is found to slowly move outwards after the bounce, eventually reaching the horizon at which time the horizon goes away and there is no black hole anymore; the lifetime of the black hole (between the initial formation of the horizon and its disappearance when the shock exits it) was found to be $\sim M^2 / m_{\rm Pl}$. As a first step in checking the robustness of these results, it is necessary to relax the assumption of the dust field being marginally bound. It is especially important to ensure that the formation of a shock is a robust prediction, since there were no hints of such an effect in earlier work---in part this is not entirely surprising since shocks can only form when there are local degrees of freedom, so models with a finite number of degrees of freedom will necessarily be blind to the possibility of a shock forming. On the other hand, it has recently been shown that it is possible to obtain a model for Oppenheimer-Snyder collapse in LQG without the formation of a shock \cite{Fazzini:2023scu, Giesel:2023hys} (although at the expense of either a discontinuity in the dust field or a non-monotonic areal radius). However, it turns out that the Oppenheimer-Snyder model is finely tuned and its dynamics are not representative: at least in the marginally bound case, for any collapsing profile of dust where the dust energy density is continuous and of compact support, a shell-crossing singularity will necessarily form \cite{Fazzini:2023ova}, signaling the formation of a shock wave and the need to consider weak solutions to the dynamics. This most recent result suggests that a shock will form in typical solutions to dust collapse in LQG, at least in the marginally bound case. By studying configurations beyond this special case, it will be possible to determine how generally a shock forms during dust collapse in LQG.

Further, the studies that have considered gravitational collapse in LQG with non-vanishing spatial curvature have found that, at least in the simple case of the Oppenheimer-Snyder model, in the presence of positive spatial curvature the dynamics become cyclical: the collapse phase ends at a bounce (generated by LQG effects) and then the radius of the star begins to increase, but due to the positive spatial curvature, the interior is (a portion of) a closed Friedman universe which will eventually recollapse, which will be followed by a bounce, and so on. In this way, the result is a cyclical process that has been described as a `pulsating star' \cite{BenAchour:2020mgu, Cafaro:2024}. Since these results have so far only been obtained in models for gravitational collapse that are restricted to a finite number of degrees of freedom, it is of interest to determine whether such cyclical dynamics also arise when there are local degrees of freedom.  In particular, if a shock forms, how does this modify the picture? Does the shock move cyclically, or are the dynamics no longer cyclic? And although this discussion has focused on the case of positive spatial curvature, it is also interesting to understand the impact of negative spatial curvature on the dynamics as well. By finding numerical solutions to the dynamics, it will be possible to answer these questions, and determine the role of spatial curvature (whether positive or negative) on gravitational collapse in LQG.

The outline of the paper is as follows. We start by briefly reviewing the LQG effective dynamics for LTB spacetimes in Sec.~\ref{s.ltb}, and in Sec.~\ref{s.cosm} we take the homogeneous limit (for arbitrary spatial curvature), to obtain the Friedman universe and recover some results of loop quantum cosmology. Then, in Sec.~\ref{s.sharp} we study configurations corresponding to a star with a sharp boundary, close to the Oppenheimer-Snyder model (with non-vanishing spatial curvature) but with a continuous energy density for the dust. It is possible to approximately solve the dynamics for the collapse for such a configuration, with the result that there is a bounce, and a shock forms very soon after the bounce and moves outwards. In this section we also discuss the properties of the vacuum exterior, which are in many respects similar to what was found for the spatially flat case, but with the difference that there is quantum hair due to the presence of spatial curvature. Finally, in Sec.~\ref{numerical_section} we solve the dynamics numerically for a wide range of initial conditions, first confirming the results found for stars with sharp boundaries obtained in the previous section, and then considering other types of configurations. The general picture is always the same: during the collapse, a bounce occurs in the Planck regime due to LQG effects, and a shock wave forms at the latest very soon after the bounce. We end with a short discussion in Sec.~\ref{s.con}.

We use units where $c=1$ throughout; in Sec.~\ref{numerical_section} devoted to the numerical analysis, we additionally use units where $G = \hbar = 1$ and set the Barbero-Immirzi parameter to $\gamma = 1$.

\section{Effective LQG dynamics for LTB spacetimes}
\label{s.ltb}

In generalized Painlevé-Gullstrand (GPG) coordinates, the metric for the LTB spacetime is
\begin{equation} \label{metricpg}
 \dd s^{2}=-\dd t^{2}+\frac{1}{1+\varepsilon(x,t)}(\dd x+N^{x}\dd t)^{2}+x^{2}\dd \Omega^{2},
\end{equation} 
where $\dd \Omega^2 = \dd \theta^2 + \sin^2 \theta \, \dd\phi^2$ and $\varepsilon > -1$.

LQG effects on the dynamics of LTB spacetimes can be captured through a set of effective equations of motion where corrections (proportional to $\hbar$) modify the classical Hamiltonian, expressed in terms of connection and triad variables, generating the dynamics.  These corrections are due to the fundamental discreteness of quantum geometry predicted by LQG and their main effect is to effectively cause a repulsive force when the spacetime curvature nears the Planck scale.

Different approaches have been proposed to derive the LQG effective dynamics for LTB spacetimes \cite{Bojowald:2008ja, Kelly:2020lec, Alonso-Bardaji:2020rxb, Giesel:2021dug, Husain:2021ojz, Husain:2022gwp, Giesel:2022rxi, Giesel:2023tsj, Alonso-Bardaji:2023qgu}, here we use the effective dynamics derived by performing the so-called `K' loop quantization after imposing the areal and dust-time gauges before quantization \cite{Kelly:2020lec, Husain:2021ojz, Husain:2022gwp}; it has recently been shown that the same effective equations can also be obtained without needing to impose the areal gauge before the loop quantization \cite{Giesel:2023tsj, Giesel:2023hys} (and these dynamics also follow from a covariant action based on a mimetic theory of modified gravity \cite{Giesel:2023hys}).  In this effective theory, following the notation in \cite{Kelly:2020uwj, Kelly:2020lec},
the shift vector takes the form
\begin{equation}
N^{x}=-\frac{x}{\gamma \sqrt{\Delta}}\sin\frac{\sqrt{\Delta}b}{x}\cos\frac{\sqrt{\Delta}b}{x},
\end{equation}
where $b$ is the component of the extrinsic curvature in the angular direction (for further details, see \cite{Kelly:2020uwj}), $\gamma$ is the Barbero-Immirzi parameter and $\Delta$ is the minimum area gap in loop quantum gravity $\Delta \sim \lp^2$; the effective dynamics are generated by the Hamiltonian
\begin{equation}
 \mathcal{H}_{phys}=-\frac{1}{2 G\gamma}\left[\frac{|E^{b}|}{\gamma \Delta x}\partial_{x}\left(x^{3}\sin^{2} \frac{\sqrt{\Delta }b}{x} \right)+\frac{\gamma |E^{b}|}{x}+\frac{\gamma x}{|E^{b}|} \right].
\end{equation}
Here, $E^b$ is the component of the densitized triad in the angular directions related to metric components through $(E^b)^2 = x^2/(\varepsilon + 1)$, and is canonically conjugate to $b$: $\{ b(x_1), E^b(x_2) \} = G \gamma \, \delta(x_1 - x_2)$. 
Note that $\mathcal{H}_{phys}$ is derived from a gauge-fixing of the Hamiltonian constraint by using the dust field as a relational clock \cite{Kelly:2020lec}, as a result $\mathcal{H}_{phys}$ is the physical Hamiltonian (not a constraint), and does not necessarily vanish. 
The equations of motion can be derived from the physical Hamiltonian; assuming $E^{b}>0$,
\begin{align}\label{eq:semiclassical_eqs}
\dot{E}^b &=\{ E^{b},\mathcal{H}_{phys}\}   = -\frac{x^2}{\gamma \sqrt{\Delta}}\partial_x\left(\frac{E^b}{x}\right)\sin\frac{\sqrt{\Delta}b }{x}\cos\frac{\sqrt{\Delta}b }{x},\\
\dot{b} &= \{b,\mathcal{H}_{phys}\}  = \frac{\gamma}{2}\left( \frac{x}{\left(E^b\right)^2} - \frac{1}{x} \right) - \frac{1}{2\gamma\Delta x}\partial_x \left( x^3 \sin^2\frac{\sqrt{\Delta}b }{x} \right), \label{eqsemi2}
\end{align}
and the energy density of the dust field is given by
\begin{equation}
    \rho=-\frac{\mathcal{H}_{phys}}{4 \pi x |E^{b}|}= \frac{1}{8 \pi G x^{2}}\partial_{x} \left[ \frac{x^{3}}{\gamma^{2}\Delta}\sin^{2} \frac{\sqrt{\Delta}b}{x} - \frac{x^{3}}{(E^{b})^{2}}+x  \right].
    \label{rhoH}
\end{equation}

The two equations \eqref{eq:semiclassical_eqs} and \eqref{eqsemi2} can be simplified by changing variables from $E^b$ to $\varepsilon$,
\begin{equation}
 \varepsilon=\frac{x^{2}}{(E^{b})^{2}}-1 .
 \label{var}
\end{equation}
with the result
\begin{align}
        &\dot{\varepsilon} = -\frac{x}{\gamma \sqrt{\Delta}}(\partial_x \varepsilon)\sin\frac{\sqrt{\Delta}b }{x}\cos\frac{\sqrt{\Delta}b }{x},
        \label{set0}\\
&\dot{b} = \frac{\gamma}{2x}\varepsilon - \frac{1}{2\gamma\Delta x}\partial_x \left( x^3 \sin^2\frac{\sqrt{\Delta}b }{x} \right)  ,
\label{set}
\end{align}
and the dust energy density becomes
\begin{equation}
\rho= \frac{1}{8 \pi G x^{2}} \, \partial_{x} \left[ \frac{x^{3}}{\gamma^{2}\Delta}\sin^{2} \frac{\sqrt{\Delta}b}{x} - x \varepsilon  \right].
\end{equation}
The family of solutions with $\varepsilon=0$ (known as marginally bound) has already been studied in some detail \cite{Husain:2021ojz, Husain:2022gwp, Fazzini:2023ova}; the dynamics simplify considerably since \eqref{set0} is automatically satisfied, and for this set of solutions the equation for $b$ becomes a conservation law (after rescaling $b$ by a factor of $x$). Here we will focus on the case $\varepsilon \neq 0$, which requires solving two coupled non-linear partial differential equations that are not conservation equations, rendering them challenging to handle whether analytically or numerically.  In the following, we first consider some particularly simple configurations where analytical solutions can be derived (either exact or approximate, depending on the case), and then develop and use numerical tools tailored to solve these equations of motion.

\section{Homogeneous and Isotropic Cosmology}
\label{s.cosm}

As a first step, it is interesting to consider the homogeneous limit of the LTB spacetime, corresponding to the Friedman-Lema\^itre-Robertson-Walker (FLRW) spacetimes.

In the standard comoving Friedman coordinates, the line element (for arbitrary spatial curvature $k$) for the FLRW spacetime is
\begin{equation}
\dd s^{2}=-\dd t^{2}+a(t)^{2}\left(\frac{\dd r^{2}}{1-kr^{2}}+r^{2}\dd \Omega^{2} \right).
\end{equation}
A connection with the line element \eqref{metricpg} for the LTB spacetime in GPG coordinates is obtained by the change of coordinates $x(t,r)=a(t) \cdot r$, with a brief calculation giving
\begin{equation}
 \dd s^{2}=-\left(1-\frac{ H^2 x^{2}}{1-\frac{kx^{2}}{a(t)^{2}}} \right)\dd t^{2}-\frac{2 H x}{1-\frac{kx^{2}}{a(t)^{2}}} \, \dd t \, \dd x+\frac{1}{1-\frac{kx^{2}}{a(t)^{2}}}\dd x^{2}+x^{2}\dd \Omega^{2}   ;
\end{equation}
where $H = \dot{a} /a$ is the Hubble rate.
A direct comparison between the two line elements indicates that
\begin{equation} \label{nx}
N^{x}=-H x=-\frac{x}{\gamma \sqrt{\Delta}} \sin \frac{\sqrt{\Delta}b}{x}\cos\frac{\sqrt{\Delta}b}{x}, \qquad \quad
\varepsilon = -k \cdot \frac{x^{2}}{a^{2}}.
\end{equation}

The system of equations \eqref{eq:semiclassical_eqs} and \eqref{eqsemi2} can be rewritten in terms of $a(t)$ and $\rho(t)$. Squaring the first relation of \eqref{nx},
\begin{equation}
H^2 = \f{1}{\gamma^2 \Delta} \sin^{2}\left(\frac{\sqrt{\Delta}b}{x} \right)\left( 1-\sin^{2} \frac{\sqrt{\Delta}b}{x} \right),
\end{equation}
and then combining \eqref{rhoH} and the second equality of \eqref{nx},
\begin{equation}
 H^2 = \left( \frac{8\pi G}{3} \rho -\frac{k}{a^{2}}\right) \left( 1-\frac{\rho}{\rho_{c}} +\frac{3k}{8 \pi G \rho_c a^{2}} \right),
 \label{efffried}
\end{equation}
where $\rho_{c}\equiv 3 / (8 \pi G \gamma^{2}\Delta)$ is the critical energy density in LQC; this is in agreement with \cite{Cafaro:2024}, and is precisely the LQC effective Friedman equation, for any spatial curvature $k$, following the `K' loop quantization derived in earlier work focused on homogeneous and isotropic cosmology \cite{Vandersloot:2006ws, Singh:2013ava} (the effective Friedman equation for the spatially flat $k=0$ case was previously derived from the LQG effective dynamics for LTB spacetimes in \cite{Kelly:2020lec}).

The continuity equation can be derived by differentiating \eqref{rhoH} with respect to $t$ and using both relations in \eqref{nx}, giving
\begin{equation}
\Dot{\rho}= \frac{H}{4 \pi G\gamma x^2} \, \partial_x \left( x^2 \Dot{b} - \frac{\gamma k x^3}{a^{2}} \right)=-3H \rho,
\end{equation}
and the last equality follows from using \eqref{set} and then \eqref{rhoH}. Note that there are no quantum corrections to the continuity equation, again exactly as found in LQC. These derivations of the LQC effective Friedman and continuity equations (which can be combined to derive the Raychaudhuri equation) show the robustness of the earlier results in LQC, and provide evidence that the effective dynamics for the LTB spacetimes are capturing the same physics as what was earlier found in a simpler context.

The effects due to LQC cause a non-singular bounce to occur in the Planckian regime when the terms in the second set of parentheses on the right side of \eqref{efffried} vanish. Away from the bounce, LQG effects rapidly become negligible and general relativity becomes an excellent approximation (the classical Friedman equation can be obtained in the limit $\Delta \to 0$, which sends $\rho_c \to \infty$). There will be a single bounce in the spatially open and spatially flat cases ($k=0,-1$), and an infinite number of bounces for the spatially closed case ($k=+1$) due to the recollapses that occur due to the spatial curvature.  For further details on the LQC of FLRW spacetimes, see, e.g., the review \cite{Ashtekar:2011ni}.

\section{Collapse of a star with a sharp boundary}
\label{s.sharp}

The simplest model for gravitational collapse is the Oppenheimer-Snyder configuration, where the interior is composed of a homogeneous dust field and the exterior is vacuum.  While this simple model is of considerable interest since it is possible to find exact analytical solutions, in the context of the LQG effective dynamics it is unusual in that a shell-crossing singularity does not form, as is discussed in more detail in App.~\ref{appendix:SCSing}. This is different from the generic case, since (at least for $\varepsilon = 0$) all initial profiles for the dust energy density that are continuous and of compact support necessarily lead to the formation of a shell-crossing singularity at which point a shock is formed \cite{Fazzini:2023ova} . Although this last result has so far only been proven for $\varepsilon = 0$, it seems likely to be true for $\varepsilon \neq 0$ as well, and we have checked numerically that a shell-crossing singularity does indeed occur for the initial configurations (for which $\rho$ is continuous and of compact support) considered in this section, see App.~\ref{appendix:SCSing} for details.

Due to this important limitation of the Oppenheimer-Snyder model, here we will study configurations that are close to Oppenheimer-Snyder (and therefore can in some contexts be approximated by such a configuration to a high degree of accuracy), but nonetheless have a continuous initial profile for $\rho$, $\varepsilon$ and $b$ for which a shell-crossing singularity occurs, leading to the formation of a shock.

\subsection{Initial data}
\label{s.sharp-in}

A simple initial energy density profile that is continuous, with a homogeneous interior and a sharp boundary is
\begin{equation}
 \rho(t_{0})= \begin{cases}
     \rho_{0}, \quad\quad\quad\,\, & \text{for $x<x_{0}$, }    \\ \displaystyle
    \rho_{0} \cdot \frac{x_{1}-x}{x_{1}-x_{0}}, \quad & \text{for $ x_{0}<x<x_{1}$,}\\
    0, \quad\quad \quad\;\;\, &\text{for $x>x_{1}$,}
 \end{cases}   
 \label{rhocont}
\end{equation}
here $x_0$ is the boundary of the inner homogeneous region, and $\rho(x,t_0)$ decreases linearly between $x_0$ and $x_1$ where it reaches 0.
By taking $x_{1}-x_{0} \ll x_0$, the boundary to the star can be made arbitrarily sharp. Up to the bounce, the dynamics of such a configuration---as confirmed numerically---can be approximated by the Oppenheimer-Snyder configuration obtained as the limiting case $x_{1}\rightarrow x_{0}$ wherein
\begin{equation} \label{rhoo}
\rho(t_{0}) \to \rho_{OS}(t_0) = \rho_0 \Big[1-\theta\left(x-x_{0}\right)\Big],
\end{equation}
in this limit $x_{0}$ is the areal radius of the Oppenheimer-Snyder star at the initial time $t_{in}=t_{0}$, and $\theta$ is the Heaviside function so $\rho_{OS}(t_0)$ vanishes for $x > x_0$.

Since the innermost region is a portion of an FLRW universe,
\begin{equation}
\varepsilon_{in}(t_{0},x)=-k\frac{x^{2}}{a(t_{0})^{2}}.
\label{lasagna}
\end{equation}
For the exterior, we also wish to choose a configuration close to Oppenheimer-Snyder. For the Oppenheimer-Snyder model, the exterior is determined by enforcing the Israel junction conditions \cite{Poisson:2009}, and it is a straightforward (although tedious) calculation to show that in the exterior region $\varepsilon = -k x_{0}^{2} / a(t_{0})^2$. To mimic this behaviour, we set the continuous initial condition for $\varepsilon$  
\begin{equation}
 \varepsilon(t_{0},x)= \begin{cases}
    \displaystyle -\alpha  \cdot\frac{x^{2}}{x_{0}^{2}}, \quad\quad\quad\,\, & \text{for $x<x_{0}$, }    \\
    \displaystyle -\alpha, \quad & \text{for $ x > x_{0}$,}
 \end{cases}   
 \label{epsilon}
\end{equation}
where $\alpha = k x_0^2 / a(t_0)^2$.

Inverting relation \eqref{rhoH},
\begin{equation}\label{eq:IDb}
 b(t_0,x) = -\frac{x}{\sqrt{\Delta}}\sin^{-1} \Bigg[ \sqrt{ \frac{\gamma^2 \Delta}{x^3} \left( 8 \pi G \int_{0}^{x} \!\! \dd \tilde x \: \tilde x^{2}\rho(\tilde x) + x \varepsilon \right) } \: \Bigg],
\end{equation}
where the negative root is chosen so the profile is initially collapsing rather than expanding. Therefore, $b$ is fully determined (up to an overall sign corresponding to contraction or expansion) by the choice of initial conditions for $\rho$ and $\varepsilon$; for the choices made above,
\begin{equation}
\!\!    b(t_0,x)= \begin{cases}
        -\frac{x}{\sqrt{\Delta}}\sin^{-1}\left[\sqrt{\frac{\gamma^{2}\Delta}{x^{3}}\left( \frac{8}{3}\pi G \rho_{0}x^{3} -\frac{\alpha x^3}{x_{0}^{2}}        \right)    }   \right], & \text{for} ~ 0<x<x_{0},     \\
        -\frac{x}{\sqrt{\Delta}}\sin^{-1}\left[\sqrt{\frac{\gamma^{2}\Delta}{x^{3}}\left( \frac{2}{3}\pi G \rho_{0} \cdot \frac{4x_{1}x^{3}-3x^{4}-x_{0}^{4}}{x_{0}-x_1} -x \alpha       \right)    }   \right]  , & \text{for} ~ x_{0}<x<x_{1},   \\
        -\frac{x}{\sqrt{\Delta}}\sin^{-1}\left[\sqrt{\frac{\gamma^{2}\Delta}{x^{3}}\left( \frac{2}{3}\pi G \rho_{0}(x_{0}+x_{1})(x_{0}^{2}+x_{1}^{2})  -x \alpha       \right)    }   \right],  & \text{for} ~ x>x_{1}.
        \end{cases}
\end{equation}

\subsection{Vacuum exterior solutions}
\label{subsub:vacuum}

Before studying the dynamics, it is worth reviewing in some detail the general vacuum exterior solution for $x > x_1$ obtained from the effective LQG dynamics for the LTB spacetime considered here; this has also been studied in \cite{Cafaro:2024}. For a more general discussion of the effective geometry of vacuum spherically symmetric spacetimes in LQG, including a review of different models that have been proposed in the literature, see the recent reviews \cite{Perez:2017cmj, Gambini:2022hxr, Ashtekar:2023cod}.

The stationary solutions to \eqref{set0}--\eqref{set}, putting aside solutions for which $\sin \frac{\sqrt{\Delta}b}{x} \cos \frac{\sqrt{\Delta}b}{x} =0$ (where $\sin \f{\sqrt\Delta b}{x} = 0$ corresponds to the Minkowski spacetime, and $\cos \f{\sqrt\Delta b}{x} = 0$ corresponds to a fully quantum solution to the effective dynamics with no classical equivalent that requires a negative energy density $\rho < -\rho_c$), satisfy the conditions
\begin{equation}
\partial_{x}\varepsilon=0, \qquad \qquad
\varepsilon = \frac{1}{\gamma^{2}\Delta}\partial_{x}\left[x^{3}\left( \sin^{2}\frac{\sqrt{\Delta}b}{x} \right)   \right].
\end{equation}
The first condition shows that for these stationary solutions, $\varepsilon$ is independent of $t$ and $x$.

The simplest such solution is $\varepsilon=0$, which has already been analyzed in \cite{Kelly:2020uwj}, but other constant values are also possible,
\begin{equation}
   \varepsilon= -\alpha,
   \label{alpa} 
\end{equation}
where the minus sign is included for later convenience, and $\alpha<1$ as required by \eqref{metricpg}.

Substituting this expression into the second equation of the system and integrating over $x$,
\begin{equation}
   b=-\frac{x}{\sqrt{\Delta}}\sin^{-1}\left(\sqrt{ -\frac{\gamma^{2}\Delta \alpha}{x^{2}}+ \frac{\gamma^{2}\Delta C}{x^{3}}  } \right) ,
   \label{bfield}
\end{equation}
and to obtain the correct classical limit the integration constant C is fixed to $R_S = 2GM$, where as usual $M$ is the gravitational mass of the interior region. Then, the shift vector $N^{x}$ is
\begin{equation} \label{nx-alpha}
N^{x}= \sqrt{-\alpha+ \frac{R_S}{x}} \cdot \sqrt{1+\frac{\gamma^{2}\Delta \alpha}{x^{2}}-\frac{\gamma^{2}\Delta R_S}{x^{3}}}.
\end{equation}
These vacuum solutions are to some extent the effective counterpart of Martel-Poisson coordinates \cite{Martel:2000rn} although (as shall be discussed below) they are not diffeomorphic to the solution with $\varepsilon=0$ as is the case in general relativity. The resulting line element is
\begin{equation}
 ds^{2}=-\dd t^{2}+\frac{1}{1-\alpha}\left(\dd x+\sqrt{-\alpha+ \frac{R_S}{x}} \cdot\sqrt{1+\frac{\gamma^{2}\Delta \alpha}{x^{2}}-\frac{\gamma^{2}\Delta R_S}{x^{3}}}\dd t \right)^{2}+x^{2}\dd \Omega^{2},    
\end{equation}
in agreement with \cite{Cafaro:2024}. It is straightforward to verify that $\rho = 0$, so this is a vacuum solution.  (As an aside, note that this calculation does not prove that there cannot be other vacuum solutions beyond these, but any such solutions, if they exist, would necessarily not be stationary, see also \cite{Cafaro:2024}.)

The expression for $N^x$ contains two square roots, which limit its domain. The argument of the first square root is always positive for $\alpha \le 0$, but for positive $\alpha$ the $N^x$ is real only for
\begin{equation}
 x \leq \frac{R_S}{\alpha},
\end{equation}
so the generalized Painlev\'e-Gullstrand coordinates used here only hold up to this maximal areal radius; other coordinates must be used beyond this point. Importantly, the generalized Painlev\'e-Gullstrand coordinate system is nonetheless valid for the central region where quantum gravity effects are strongest and that we are most interested in here.

The condition that the argument for the second square root be positive implies
\begin{equation}
 1+\frac{\gamma^{2}\Delta \alpha}{x^{2}}-\frac{\gamma^{2}\Delta R_S}{x^{3}}\geq 0 ,
 \label{xmin}
\end{equation}
establishing a lower bound $x_{min}$, that depends on $\alpha$, for the domain of the metric for the vacuum solution in these coordinates.
For the collapse models of interest here, this lower bound is not an issue, since (as we discuss below) we find that the matter cannot be compressed into a radius smaller than $x_{min}$. Rather, $\rho \neq 0$ in the region $x < x_{min}$, and we find that generalized Painlev\'e-Gullstrand coordinates can be used to describe this interior region as well (where the dust energy density is non-vanishing and therefore the metric has a different form than the vacuum solution considered in this section).

\begin{figure}
    \centering
    \includegraphics[width=\textwidth]{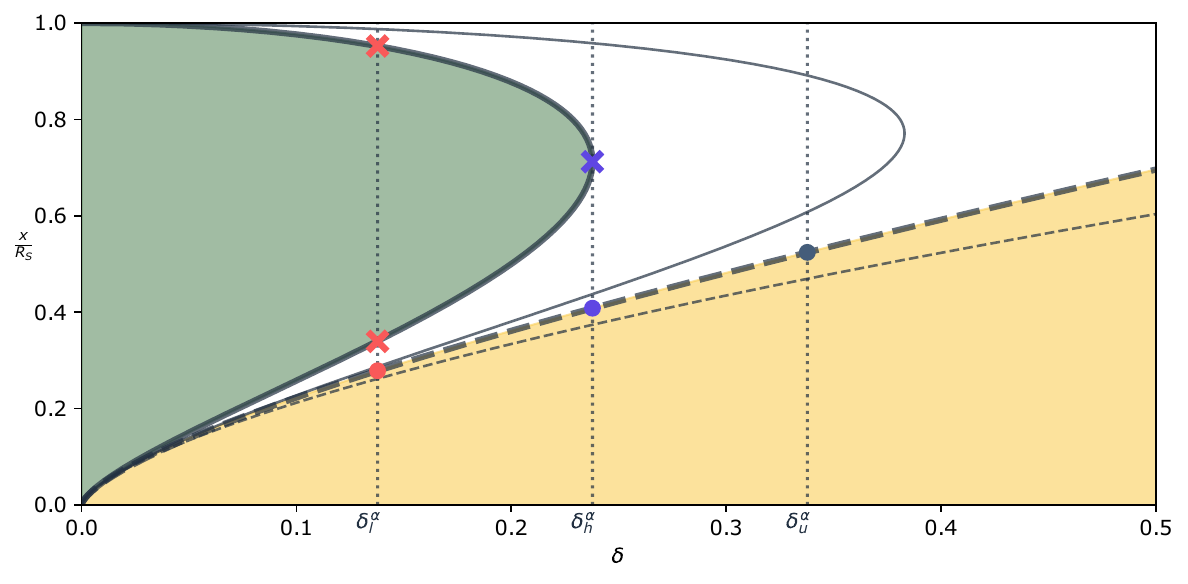}
    \caption[]{\footnotesize
    This figure shows, for $\alpha = -0.5$, the position of the inner boundary of the domain $x_{min}$ (Eq. \eqref{xmin}, thick dashed line) and the locations of the inner and outer Killing horizons $x_{in}$ and $x_{out}$ (Eq. \eqref{eq:exactHor}, thick solid line) as a function of $\delta = \gamma \sqrt{\Delta} / R_S$. The region shaded in yellow lies below $x_{min}$, while the region shaded in green shows the trapped region between the Killing horizons. For a given mass---i.e., fixed $\delta$---there is always an $x_{min}$; three values of $\delta$ are shown with dotted lines, and the corresponding $x_{min}$ marked with a circle. The location and number of horizons change with $\delta$: if $\delta<\delta^{\alpha}_h$ there are two (marked with red crosses), if $\delta=\delta^{\alpha}_h$ there is only one (marked with a violet cross); otherwise there is none. The thinner dashed and solid lines show the curves of Eqs.~\eqref{xmin} and \eqref{eq:exactHor} for $\alpha=0.2$.}
    \label{fig:deltaVSxtilde}
\end{figure}

It is convenient to introduce the dimensionless variable $\bar{x}= x / R_S$, in terms of this variable the condition \eqref{xmin} becomes
\begin{equation}
\delta \equiv \frac{\gamma \sqrt{\Delta}}{R_S}\leq \sqrt{\frac{\bar{x}^{3}}{1-\alpha \bar{x}}}.
\end{equation}
This relation is plotted in Fig.~\ref{fig:deltaVSxtilde} for two different values of $\alpha$, the regions below the dashed line lie in the region $x < x_{min}$.

Next, depending on the value of $R_{S}$ and $\alpha$, the vacuum solution can have two, one, or zero Killing horizons where the norm of the Killing field $\xi^{\mu}=(1,0,0,0)$ vanishes, which gives the relation
\begin{equation}\label{eq:exactHor}
 1-\frac{R_S}{x}+\frac{\gamma^{2}\Delta \alpha^{2}}{x^{2}}-\frac{2\gamma^{2}\Delta R_S\alpha}{x^{3}}+\frac{\gamma^{2}\Delta R_S^{2}}{x^{4}}=0,
\end{equation}
determining the location of the Killing horizons.

This equation can also be expressed in terms of $\bar x$ and $\delta$, and the locations of the Killing horizons are shown by the solid curve enclosing the green region in Fig.~\ref{fig:deltaVSxtilde}; the green region lies between the two Killing horizons and is trapped. Note that, for each $\alpha$, there is a critical value for $R_{S}$ where there is one Killing horizon, and for smaller $R_S$ there is no horizon at all. Equivalently, in terms of $\delta$, there is a critical value $\delta_h^\alpha$ (that is easily computed numerically for any value of $\alpha$) for which there is one Killing horizon, and for $\delta > \delta_h^\alpha$ there are no Killing horizons.

To make contact with the classical limit, the location of the outer horizon can be expanded in powers of $\Delta/R_S^2$,
\begin{equation} \label{outerh}
   x_{\text{outer}} = R_S \left( 1 - (\alpha-1)^2 \f{\gamma^2 \Delta}{R_S^2} + O\left(\frac{\Delta^2}{R_S^4}\right) \, \right), 
\end{equation}
and doing the same for the inner horizon gives
\begin{equation}
   x_{\text{inner}} = (\gamma^2 \Delta R_S)^{1/3} \left( 1 + \frac{1-2\alpha}{3} \left(\frac{\gamma^2 \Delta}{R_S^2}\right)^{1/3} + \f{(\alpha-1)^2}{3} \left( \frac{\gamma^2 \Delta}{R_S^2} \right)^{2/3} + O\left(\frac{\Delta}{R_S^2} \right) \, \right).
\end{equation}

It is also of interest to calculate some curvature scalars for the vacuum solution, for example
\begin{align}
R =&\, \f{\gamma^2 \Delta R_S}{x^6} \Big( 4 \alpha x - 6R_S \Big), \\ 
R_{\mu \nu}R^{\mu \nu} =&\, \f{\gamma^4 \Delta^2}{x^{12}} \Big( 4\alpha^{4}x^{4}-40R_S\alpha^{3}x^{3}+140R_S^{2}\alpha^{2}x^{2}-192R_S^{3}\alpha x+90R_S^{4} \Big).
\end{align}
As expected, these curvature scalars both vanish in the classical limit $\Delta \to 0$.  (Note also that although these expressions diverge at $x=0$, that point lies outside the domain of validity of these coordinates for the vacuum solution: the curvature scalars are finite everywhere these coordinates hold for the vacuum solution.)

The vacuum solution for $\alpha \neq 0$ is qualitatively similar to the $\alpha=0$ vacuum solution studied in \cite{Kelly:2020uwj}, but differs in the quantitative value of the curvature scalars and location of the Killing horizons; this signifies the failure of the no-hair theorem as these static solutions do not correspond to the same spacetime geometry. In classical general relativity, vacuum solutions with different values of $\alpha$ are diffeomorphic to each other: one can be transformed into the other by changing the time coordinate (but leaving the radial coordinate $x$ the same). This is not the case for the effective LQG vacuum solutions being considered here, as is most obvious from the dependence of the curvature scalars on $\alpha$. Therefore, although different values of $\alpha$ corresponded to different coordinate systems describing the same spacetime in classical general relativity, the effective LQG solutions with different values of $\alpha$ correspond to different spacetime geometries. There are more solutions to the LQG effective dynamics, which signals the presence of quantum hair: the vacuum solutions depend also on $\alpha$, in addition to $M$.

In consequence, for these effective LQG dynamics the vacuum spherically symmetric solutions have quantum hair due to the spatial curvature (for the slicing implied by using GPG coordinates) of the spacetime.

\subsection{Pre-bounce dynamics}
\label{subsub:prebounce}

Returning to the study of the dynamics of the gravitational collapse, it is convenient to split the analysis into two parts, before and after the bounce. Before the bounce, for a configuration with a sharp boundary as considered here, the dynamics can be approximated to a high degree of precision by the Oppenheimer-Snyder collapse, as can be confirmed numerically. For this reason we will solve the dynamics for the collapse phase here for the Oppenheimer-Snyder configuration \eqref{rhoo} as an approximation to the sharp boundary configuration \eqref{rhocont}; in the two cases the initial spatial curvature is the same, given by \eqref{epsilon}.

For the Oppenheimer-Snyder model for gravitational collapse, the variable of interest is the radius $L(t)$ of the idealized star. Since the interior is a portion of an FLRW spacetime, the radius $L$ is proportional to the scale factor of the FLRW spacetime,
\begin{equation}
  L(t)=a(t) \cdot r_{0},
  \label{a}
\end{equation}
where $r_0$ is a constant. Substituting into the effective Friedmann equation \eqref{efffried} derived above,
\begin{equation}
 \left(\frac{\Dot{L}}{L} \right)^{2}=\left(\frac{8\pi G}{3} \rho-\frac{k r_{0}^{2}  }{L(t)^{2}}\right)    \left(1-\frac{\rho}{\rho_{c}} +  \frac{3 k r_{0}^{2}}{8 \pi G \rho_c L(t)^{2}} \right).
 \label{effried1}
\end{equation}
To maintain consistent notation with the previous sections, we denote
\begin{equation}
\alpha=k \frac{L(t_{0})^{2}}{a(t_{0})^{2}},
\label{b}
\end{equation}
which fixes $r_{0}^{2}= \alpha/ k$. Finally, substituting this back into~\eqref{effried1},
\begin{equation}
    \left(\frac{\Dot{L}}{L} \right)^{2}=\left(\frac{8 \pi G}{3} \rho-\frac{\alpha  }{L(t)^{2}}\right)    \left(1-\frac{\rho}{\rho_{c}} + \frac{3}{8 \pi G \rho_c} \cdot \frac{\alpha}{L(t)^{2}}  \right)   ,
    \label{finall}
    \end{equation}
or, expressing $\rho$ in terms of the gravitational mass $M = 4 \pi \int_0^L \dd x \: x^2 \, \rho = 4 \pi L^3 / 3$ of the Oppenheimer-Snyder star,
\begin{equation}
    \left(\frac{\Dot{L}}{L} \right)^{2}=\left(\frac{R_S}{L^{3}}-\frac{\alpha  }{L^{2}}\right)    \left[1- \gamma^2 \Delta \left( \frac{R_S}{L^{3}} -  \frac{\alpha}{L^{2}}  \right)  \right] .
    \label{finalissima}
\end{equation}
During the collapse phase before the bounce, the dynamics of the radius of the Oppenheimer-Snyder star are given by \eqref{finalissima}, and this also provides an excellent approximation to the outer radius of a star with a sharp boundary as defined in Sec.~\ref{s.sharp-in}.

\begin{figure}
    \centering
    \includegraphics[width=0.8\textwidth]{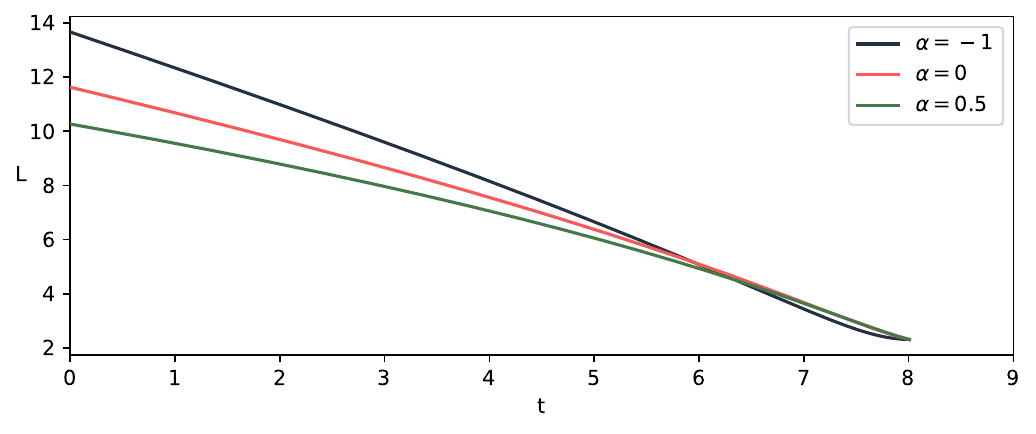}
    \caption[]{\footnotesize
    Solutions to Eq. \eqref{finalissima} for $R_S = 10$ and three different values of alpha. These solutions are valid only up to the bounce. Times and lengths are in Planck units, with $\gamma=1$ and $\Delta = 1$.}
    \label{fig:ltpreb}
\end{figure}

As shown in Fig.~\ref{fig:ltpreb}, the solution to this equation of motion is a star which collapses until it reaches a minimum value $L_{min}$ that satisfies the relation $R_S/L_{min}^3 - \alpha/L_{min}^2 = (\gamma^2 \Delta)^{-1}$, which occurs when the spacetime curvature is Planckian, and then a bounce occurs, with the star starting to move outwards---this second phase will be studied next. This general picture occurs for any value of $\alpha$, although the value of $L_{min}$ and some other quantitative results will depend on the specific value of $\alpha$. Numerical simulations show that the time between $L$ reaching $R_S$ and then reaching $L_{min}$ is of the order $\sim M$, and only weakly dependent on the value of $\alpha$.

Finally note that the minimum value allowed for $L$ from \eqref{finalissima}  is  $L_{min} = x_{min}$, so it is always possible to use generalized Painlev\'e-Gullstrand coordinates for the exterior. Further, as shown here explicitly, these coordinates can also be used for the interior during the pre-bounce collapse phase, and as we argue next, they remain valid for the interior after the bounce as well.

\subsection{Post-bounce dynamics}

A numerical analysis shows that a shell-crossing singularity forms soon after the bounce for the initial conditions considered here, see App.~\ref{appendix:SCSing} for details. This shell-crossing singularity, which as a weak singularity is not cured by effective loop quantum gravity \cite{Singh:2009mz}, indicates the formation of a shock wave \cite{Fazzini:2023ova}. (Therefore, while the Oppenheimer-Snyder dynamics provide a good approximation of a realistic collapse before and up to the bounce in effective LQG, this is no longer the case after the bounce as no shell-crossing singularity occurs for the non-generic Oppenheimer-Snyder configuration.)

When shell-crossings occur and shocks form, it is necessary to look for weak solutions to the dynamics, which solve the integral form of the equations of motion (but may not solve the differential equations if discontinuities arise dynamically in the weak solutions).
Explicitly, for an equation of the form $\dot u + \partial_x [f(u,x)] + g(u, x) = 0$, the integral form is obtained from its integral over both $x$ and $t$; the result is $\int_x u|^{t_2}_{t_1} + \int_t f(u,x)|^{x_2}_{x_1} + \int_t \int_x g(u,x) = 0$, and this is required to hold for all possible bounds for the integrals over $t$ and $x$. Although numerics (which we implement in Sec.~\ref{numerical_section}) are often necessary to gain a complete understanding of the dynamics of weak solutions, for some simple configurations analytical tools can provide important insights.

As a first step in this direction, it is convenient to rewrite the equations of motion in terms of $B = xb$, and use the inverse Liebniz rule on the equation for $\varepsilon$,
\begin{gather} \label{eq:balanceform}
  \dot{B} = - \partial_x \left( \frac{x^3}{2 \gamma \Delta} \sin^2\frac{\sqrt{\Delta}B }{x^{2}} \right) + \frac{\gamma}{2}\varepsilon, \\
  \Dot{\varepsilon}=-\partial_{x}\left[\frac{x}{2\gamma\sqrt{\Delta}} \varepsilon\sin\left(\frac{2 \sqrt{\Delta}B}{x^{2}} \right) \right]+\frac{\varepsilon}{2\gamma} \partial_{x}\left[\frac{x}{\sqrt{\Delta}} \sin\left( \frac{2 \sqrt{\Delta}B}{x^{2}}  \right) \right], \label{eq:nearbalance}
\end{gather}
and given this form of the equations of motion, it is helpful to define
\begin{equation} \label{def-mg}
m(x,B)=\frac{x^{3}}{2 \gamma \Delta} \sin^{2} \frac{B \sqrt{\Delta}}{x^{2}}, \qquad
G(x,\varepsilon,B)= \frac{x \, \varepsilon}{2 \gamma \sqrt{\Delta}} \, \sin \frac{2\sqrt{\Delta}B}{x^{2}}.
\end{equation}
In general, for a non-linear wave equation for a field $u$ of the form $\dot u = - v_u \partial_x u + J$ that potentially contains some source terms $J$ (where $J$ does not contain any derivatives of $u$), the generalized velocity of the field $u$ is given by $v_u$. In the same way, the generalized velocity of the $B$ and $\varepsilon$ fields are
\begin{equation} \label{velocities}
v_B = \partial_B m = \frac{x}{2 \gamma \sqrt{\Delta}} \, \sin \frac{2\sqrt{\Delta}B}{x^{2}}, \qquad \qquad
v_\varepsilon = \partial_\varepsilon G = v_B.
\end{equation}
The fact that the velocities of the two fields (which depend on position and the fields themselves) are identical is simply due to the fact that $B$ and $\varepsilon$ are redefinitions of the canonically conjugate variables $b, E^b$. Although expected, this result is nonetheless important in that it provides a major simplification of the dynamics and provides an avenue to solve the dynamics numerically, as shall be explained in Sec.~\ref{numerical_section}.

It is also possible to derive an equation for the velocity of the shock from the equation of motion for $B$. Since the equation for $B$ has the form of a balance law, the Rankine-Hugoniot condition \cite{Leveque:2002} can be used to calculate the velocity of the shock,
\begin{equation}
v_{\rm shock} = 
\frac{\dd L(t)}{\dd t} = \frac{[m]}{[B]}=\frac{L^{3}}{2\gamma \sqrt{\Delta}}\frac{\left.\sin^{2}\left( \frac{\sqrt{\Delta}B}{x^{2}}\right)\right|_{L^{+}} -\left.\sin^{2}\left( \frac{\sqrt{\Delta}B}{x^{2}}\right)\right|_{L^{-}}}{B(L^{+})-B(L^{-})} ,
\end{equation}
here $L(t)$ denotes the location of the shock that can be thought of as the outer boundary of the `star' during the post-bounce phase, and
\begin{equation}
 [f]=\lim_{x\rightarrow L^{+}}f(x)-\lim_{x\rightarrow L^{-}}f(x)\equiv f(L^{+})-f(L^{-}).
\end{equation}
Since field $\varepsilon$ has the same generalized velocity as $B$, the two fields will travel together in lock step.

Given this relation for the velocity of the shock, it is possible to calculate the lifetime of a black hole. The black hole is initially formed when the radius of the collapsing star lies at the location of the outer horizon $x_{\rm outer}$ (approximately equal to $R_S$). Then, the collapse ends when $L = L_{min}$ and a bounce occurs, and a shell-crossing singularity occurs shortly after the bounce signaling the formation of a shock wave near the surface of the star. As shall be shown below, the shock slowly moves outwards, eventually reaching $x_{\rm outer}$ at which time the horizon disappears and there is no longer a black hole. As mentioned in Sec.~\ref{subsub:prebounce}, the duration from the formation of a black hole to the bounce is of the order of $\sim M$, so it remains to calculate the time required for the shock to travel from $L_{min}$ to $x_{\rm outer}$, neglecting the (Planckian) time required for the shock to form after the bounce.

To evaluate this time, a certain number of assumptions are needed, even for the relatively simple case of a star with a sharp boundary; all of these assumptions are well supported by numerical results. Numerics show that after the bounce, the shock forms very near to the surface of the star, and the dust energy density at the location of the shock rapidly grows, while the energy density of the dust lying inside the shock rapidly decays so that the interior metric can soon be approximated as Minkowski. Since the shock separates an interior region that has bounced where $\sqrt{\Delta}B_{int} / x^{2} < - \pi/2$, and an exterior region that has not yet bounced where $\sqrt{\Delta}B_{ext} / x^{2} > -\pi / 2$, the approximation that the interior region tends to Minkowski gives $B_{int}(L) = - \pi x^{2} / \sqrt{\Delta}$. Further, numerics also show that the region outside the shockwave rapidly approaches the vacuum solution discussed in Sec.~\ref{subsub:vacuum}, so
\begin{equation}
 B_{ext}(L)=-\frac{L^{2}}{\sqrt{\Delta}}\sin^{-1}\left( \sqrt{  \frac{-\gamma^{2}\Delta \alpha }{L^{2}} +\frac{\gamma^{2}\Delta R_S}{L^{3}}}\right).
\end{equation}

The velocity of the shock is then
\begin{equation}
\frac{\dd L(t)}{\dd t} = \frac{L^{3}}{2 \gamma \sqrt{\Delta}}\frac{
\frac{-\gamma^{2}\Delta \alpha }{L^{2}} +\frac{\gamma^{2}\Delta R_S}{L^{3}}
}{B_{ext}(L)+\frac{\pi L^{2}}{\sqrt{\Delta}}  },
\end{equation} 
which can be simplified since soon after the bounce $L^{3} \gg L_{min}^{3}\sim \gamma^{2}\Delta R_S$, where $L_{min}$ is the minimal radius of the star, reached at the bounce, while $(\gamma^{2}\Delta R_S)^{\frac{1}{3}}$ is the location of the radius at the bounce in the spatially flat case $\alpha=0$. If we further assume that
\begin{equation}
 (\gamma^{2}\Delta R_S)^{\frac{2}{3}} \gg \gamma^{2}\Delta |\alpha|,
\end{equation}
or equivalently
\begin{equation}
R_S\gg \gamma \sqrt{\Delta} |\alpha|^{\frac{3}{2}},
\label{cond3}
\end{equation}
then $L^{2} \gg (\gamma^{2}\Delta R_S)^{\frac{2}{3}} $. This last condition holds for black holes with a mass much larger than $m_{\rm Planck}$, for which $R_S\gg\sqrt{\Delta}$, and then the inequality is satisfied for any reasonable value for $\alpha$ satisfying $\left|\alpha\right| \ll R_S / \sqrt\Delta$. 

Given these three approximations, it follows that $|B_{ext}(L)|\ll \pi L^{2} /\sqrt{\Delta}$, and
\begin{equation}
\frac{\dd L}{\dd t} \approx
\frac{\gamma \sqrt{\Delta}   }{2 \pi L^{2}}(R_S-\alpha L).
\label{velo}
\end{equation}
The time required for the shock wave to travel from $L_{min}$ to $x_{\rm outer} \approx R_S$ is given by
\begin{equation}
T_+ = \int_{L_{min}}^{R_S} \!\!  \dd L ~ \left( \frac{\dd L}{\dd t} \right)^{-1},
\end{equation}
and given the approximations above, together with the further approximation of taking the lower bound for the integral to be 0 (which gives an error only of the order of $t_{\rm Pl}$), the integral simplifies to
\begin{equation}
T_+ \approx \f{2\pi}{\gamma \sqrt\Delta} \int_0^{R_S} \! \dd L ~ \frac{L^{2}}{R_S-\alpha L},
\end{equation}
with the result
\begin{equation}\label{eq:timeTotal}
 T_+ \approx \frac{\pi R_S^2}{\gamma\sqrt{\Delta} (-\alpha)^{3}} \left[2\ln(1-\alpha) + \alpha(\alpha+2) \right],
\end{equation}
which scales as $M^2$.
This expression is positive for all values of $\alpha$.

The total lifetime of the black hole $T$ is given by the sum of the collapse time (the time elapsed between the star's radius reaching its Schwarzschild radius and reaching its minimal value $L_{min}$), the time between the bounce and the formation of the shock, and the time for the outgoing shock to reach $R_S$ given by $T_+$. In Planck units the collapse time, as discussed at the end of Sec.~\ref{subsub:prebounce}, is of the order of $M$, while the time between the bounce and the formation of the shock is of the order $t_{\rm Pl}$, so for large black holes with $M \gg m_{\rm Pl}$, the dominant contribution of order $M^2$ comes from $T_+$,
\begin{equation} \label{an-life}
T \approx T_+ \approx \frac{\pi R_S^2}{\gamma\sqrt{\Delta} (-\alpha)^{3}} \left[ 2\ln(1-\alpha) + \alpha(\alpha+2) \right].
\end{equation}
Interestingly, numerics suggest that this result also holds to a good approximation for a wide range of initial profiles for the dust energy density, beyond the configurations with a sharp boundary that have been considered in this calculation.

In the limit $\alpha \rightarrow 0$,
\begin{equation}
T \approx \frac{2\pi R_S^{2}}{3\sqrt{\Delta}\gamma},   
\end{equation}
which is precisely the result for the spatially flat case $\alpha=0$ \cite{Husain:2022gwp}.

For any fixed $\alpha$, $T \propto M^2$ exactly as was found for the spatially flat case; however, $T$ depends quite strongly on the value of $\alpha$---this is in contrast to other quantities like the location of the outer Killing horizon that only weakly depend on $\alpha$.

Finally, note that if $\alpha$ is positive, the shock will asymptotically approach a maximal radius. This can be found by integrating \eqref{velo} from $L=0$ to some $L(t)$ to solve for $t$,
\begin{equation}\label{maxL}
t = -\frac{2\pi}{\gamma \sqrt{\Delta}\alpha^{2}}\left(\frac{\alpha L^{2}}{2} + L(t)R_S + \frac{R_S^{2}}{\alpha} \ln\left(\frac{R_S - \alpha L}{R_{S}}\right) \right).
\end{equation}
From this expression, it is clear that $t$ diverges as $L\rightarrow R_S/\alpha$.

This result is starkly different from expectations obtained by considering the Oppenheimer-Snyder model (that implicitly neglects the possibility that a shock wave could form), where a cyclic `pulsating star' model was found \cite{BenAchour:2020mgu, Cafaro:2024}. In contrast, the formation of the shock changes the dynamics significantly, with the interior region rapidly emptying (where $\rho$ and $\varepsilon$ both tend to zero with increasing $t$ after the bounce) and the shock moving outwards, slowing down, and eventually asymptotically approaching a maximal radius of $R_S / \alpha$, but never recollapsing. A caveat to this result is that the calculation does depend on some assumptions, most notably that the energy density in the interior can be neglected, but this seems reasonable as the shock wave moves outwards at a slow rate, and rapidly absorbs the dust energy density of the interior which very quickly becomes highly diluted.

\section{Numerical analysis}
\label{numerical_section}

The equations of motion \eqref{eq:balanceform}--\eqref{eq:nearbalance} constitute a system of non-linear coupled partial differential equations. Due to the formation of a shock in the gravitational field, it is necessary to allow for weak solutions to the dynamics and (except for some particularly simple configurations like the case of a star with a sharp boundary considered in Sec.~\ref{s.sharp}) numerics are typically needed to solve the dynamics.

\subsection{Numerical methods}
\label{s.num}

The equations of motion can be rewritten in terms of dimensionless quantities
\begin{equation}
    x \rightarrow \sqrt{\Delta}\Tilde{x}, \quad t\rightarrow \sqrt{\Delta}\Tilde{t}, \quad B \rightarrow \sqrt{\Delta}\Tilde{B},
\end{equation}
where the tildes will be suppressed to keep the notation as simple as possible. We further use units such that $c = G = \hbar = 1$, and fix $\gamma=1$.

We discretize the fields in the radial direction on a set of points $x_j$ where $j=1, \ldots,N$, with a constant spacing $\delta x$, and in time at points $t_n$, $n=0,\ldots, M$ with a (potentially variable) spacing $\delta t$. To allow for weak solutions, we consider the integrated (in space) version of the two equations over a single spatial cell $I_j$, extending from $x_{j-1/2}=x_{j} - \frac{1}{2}\delta x$ to $x_{j+1/2}=x_{j} + \frac{1}{2}\delta x$. This approximation entails representing each field in a cell $I_j$ by its mean value:
\begin{equation} \label{discr}
     B_j(t) = \frac{1}{\delta x}\int_{x_{j-1/2}}^{x_{j+1/2}} \!\!\!\!\!\!\!\! B(t, x)~\dd x , \qquad \qquad
     \varepsilon_j(t) = \frac{1}{\delta x}\int_{x_{j-1/2}}^{x_{j+1/2}} \!\!\!\!\!\!\!\! \varepsilon(x, t)~\dd x.
\end{equation}
With these definitions, integrating the equations of motion over $I_j$ is mostly straightforward, except for the second term in the equation \eqref{eq:nearbalance} for $\dot \varepsilon$ that when integrated becomes
\begin{equation}\label{eq:approxInt}
    \frac{1}{2 \gamma} \int_{x_{j-1/2}}^{x_{j+1/2}} \!\!\!\!\!\!\!\! \dd x ~ \varepsilon \, \partial_x \left[x\sin\left(\frac{2 B}{x^2}\right)\right].
\end{equation}
To handle this term, we assume that $\varepsilon$ is nearly constant in $I_j$ and so can be approximated as $\varepsilon_j$, which can be taken outside of the integral that then contains a total derivative and so only contributes boundary terms. This approximation introduces a small numerical error that (as shall be seen below) smoothes out the shock wave in $\varepsilon$, but this error can be in large part corrected by a simple procedure described at the end of this subsection.

With this approximation, the discretized equations of motion are
\begin{gather}
    \label{eq:spatDiscrB}
    \dot{B}_j(t) = - \frac{1}{\delta x} m(x, B) \Big|^{x_{j+1/2}}_{x_{j-1/2}} + \frac{\gamma}{2}\varepsilon_j, \\
    \label{eq:spatDiscr}
    \dot{\varepsilon}_j (t) = - \frac{1}{\delta x}G(x, B, \varepsilon) \Big|^{x_{j+1/2}}_{x_{j-1/2}} +  \frac{1}{\delta x}\varepsilon_j
    \cdot v_B \Big|^{x_{j+1/2}}_{x_{j-1/2}},
\end{gather}
where $m, G$ and $v_B$ are defined in \eqref{def-mg}--\eqref{velocities}.

In general, solving two coupled non-linear wave equations can be challenging, but in this case there is an important simplification due to the fact that both fields $B, \varepsilon$ have the same speed \eqref{velocities}, and that this speed is independent of $\varepsilon$, depending only on $x, B$. As a result, it is possible to solve the equation \eqref{eq:spatDiscrB} for $\dot B_j$ first, and use the fact that the $\varepsilon$ field travels identically to solve for $\dot \varepsilon_j$ next.

To solve for $\dot B_j$, it is necessary to determine the value of $m(x, B(x))$ at the edges $x_{j\pm 1/2}$ of the cells. There are many algorithms that can reconstruct those values, we opted to employ the third-order weighted essentially non-oscillatory (WENO) method \cite{Liu:1994}. WENO is designed to minimize numerical oscillations near discontinuities in the solution. It achieves this by combining multiple low-order approximations of the field in a weighted manner, favoring smoother regions over sharp transitions. The third-order WENO algorithm proceeds as follows (for further details, see \cite{Liu:1994}):
\begin{figure}
    \centering
    \includegraphics[width=0.8\textwidth]{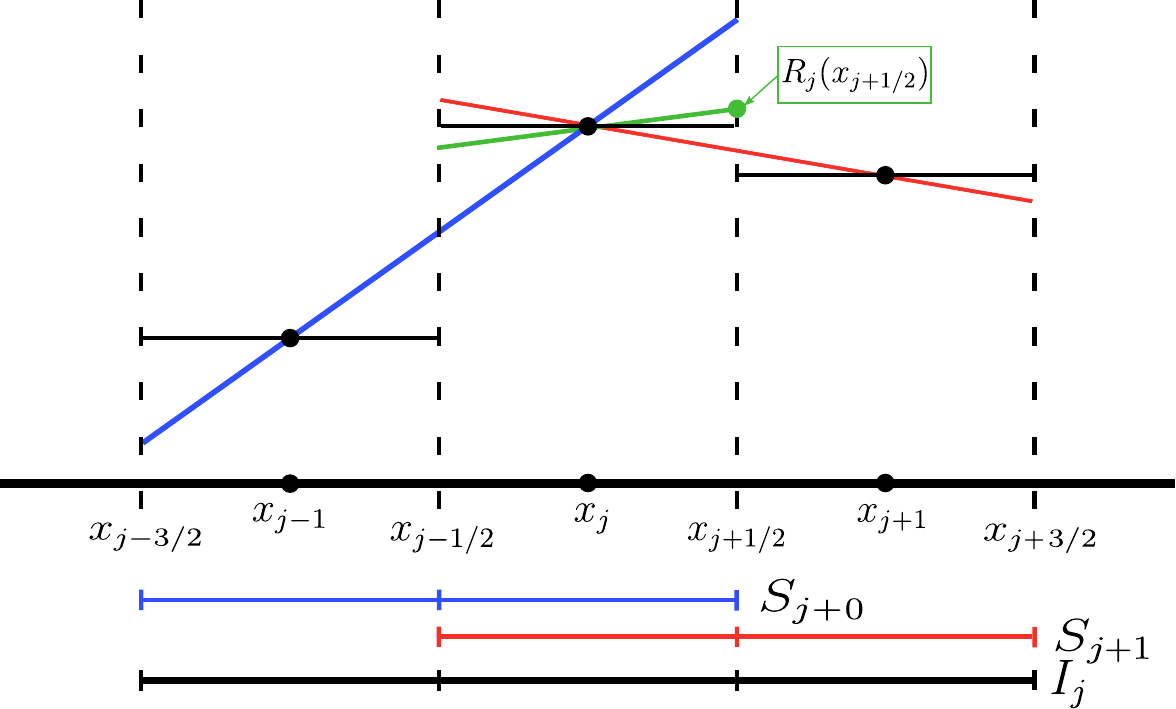}
    \caption[]{\footnotesize
    Schematic representation of the steps of the WENO algorithm for the $j$-th cell. The black lines give the initial values of the field in each cell, while the blue and red lines represent the polynomials of Eq.~\eqref{eq:interpPol}, the green line shows the linear combination given in Eq.~\eqref{eq:interpCom} and the green dot is the reconstructed value for the field at $R_j(x_{j+1/2})$, as calculated from the cell to the left of $x_{j+1/2}$.}
    \label{fig:wenoSketch}
\end{figure}

\begin{enumerate}
    \item For each cell $I_j$ define $2$ stencils $\{S_{j+k}\}_{k=0}^{1} = \{ x_{j+k-3/2},x_{j+k-1/2}, x_{j+k+1/2} \}$ composed of three points each. Then, for each $S_{j+k}$ define a linearly interpolating polynomial
    \begin{equation}\label{eq:interpPol}
    	p'_{j+k}(x) = B_{j+k-1} + \frac{B_{j+k} - B_{j+k-1}}{\delta x} (x-x_{j+k-1}),
    \end{equation}
	which captures $B$ up to $O\big( (\delta x)^2 \big)$ for the points in $S_{j+k}$.
    \item For each cell $I_j$ compute the convex combination of the polynomials for the two stencils,
    \begin{equation}\label{eq:interpCom}
        R_j(x) = \sum_{k=0}^{1} \frac{{\omega^j_k}}{\sum_{l=0}^{1} {\omega^j_l}} p'_{j+k}(x).
    \end{equation}
    The parameters $\omega^j_k$ are chosen to favour stencils where the interpolating polynomial is smoother, we follow \cite{Liu:1994} in taking (other proposals can be found in \cite{Henrick:2005, Borges:2008, Castro:2011})
    \begin{equation}
    \omega^j_0 = \frac{1}{2(\epsilon + IS_j)^2}, \quad \omega^j_1 = \frac{1}{(\epsilon+IS_{j+1})^2},
    \end{equation}
    if $v_B(x_j, B_j) \ge 0$; if instead $v_B(x_j, B_j) < 0$ the coefficients are swapped. Here $IS_j = (B_j - B_{j-1})^2$ represents a smoothness indicator and $\epsilon\sim 10^{-5}$ is a parameter included to avoid divisions by zero. This choice, away from sonic points, also boosts the accuracy of the solution to third-order \cite{Liu:1994}. 
    \item At each interface $x_{j+1/2}$ there are two reconstructed values for $B$, calculated from each adjacent cell: $R_j(x_{j+1/2})$ and $R_{j+1}(x_{j+1/2})$. To determine the value of $B$ to use in evaluating $m$ at the interface $x_{j+1/2}$ for a given time step, it is necessary to determine whether the field $B$ is moving to the right, to the left, or if there is a rarefaction wave; for the equation of motion \eqref{eq:spatDiscrB} this can be calculated from the Godunov flux:
    \begin{equation}\label{eq:Lmflux}
        m_G(x_{j+1/2}, B(x_{j+1/2})) = \begin{cases}
            \min_{R_j \leq B \leq R_{j+1}} m(x_{j+1/2}, B) \quad &\text{if } R_j \leq R_{j+1},\\
            \max_{R_{j+1} \leq B \leq R_j} m(x_{j+1/2}, B) \quad &\text{if } R_{j+1} < R_j.
        \end{cases}
    \end{equation}
    For a general discussion on the Godunov flux, see \cite{Leveque:2002}, and for more on its use to determine the dynamics of the $B$ in LQG black hole collapse models, see \cite{Husain:2022gwp}.

    Note that the flux \eqref{eq:Lmflux} can be simplified using the analytical properties of $m(x, B)$. When computing the minimum or maximum on any boundary $x_{j+1/2}$, since $B\in\left[ -\pi x_{j+1/2}^2, 0\right]$, it follows that the minimum for $m$ is always one of the endpoints while the maximum for $m$ is either one of the endpoints or the stationary point $x_{j+1/2}^3 / 2 \gamma^2$ obtained for $B = \pi x_{j+1/2}^2 / 2$.
\end{enumerate}

These three steps form the third-order WENO-Godunov algorithm to evaluate $m$ at the boundaries of the cells; 
a schematic representation of the algorithm is shown in Fig.~\ref{fig:wenoSketch}.
Following this procedure, it is possible to compute the right side of \eqref{eq:spatDiscrB}.

To calculate $\dot \varepsilon_j$, we follow the same first two steps to reconstruct two values for $\varepsilon(x_{j+1/2})$, one from each of the neighbouring cells. Then, we determine which of these two values is to be used to evaluate $G$ and $v_B$ at $x_{j \pm 1/2}$ in \eqref{eq:spatDiscr} by looking at the result of $m_G$ at that boundary point. $m_G$ could have been evaluated either (a) using the reconstructed value for $B$ coming from the cell to the left of the boundary (i.e., from $R_j(x_{j+1/2})$), (b) using the reconstructed value for $B$ coming from the cell to the right of the boundary (i.e., from $R_{j+1}(x_{j+1/2})$), or (c) using the stationary value $B = \pi x_{j+1/2}^2 / 2$. For case (a), we evaluate $G(x_{j+1/2})$ and $v_B(x_{j+1/2})$ using the $B$ and $\varepsilon$ reconstructed from the cell $I_j$, while for case (b) we evaluate $G(x_{j+1/2})$ and $v_B(x_{j+1/2})$ using the $B$ and $\varepsilon$ reconstructed from the cell $I_{j+1}$. Finally, for case (c) we use the same value of $B = \pi x_{j+1/2}^2 / 2$ in which case $G = v_B = 0$ independently of the value of $\varepsilon$.

This procedure allows one to compute the right side of \eqref{eq:spatDiscrB}--\eqref{eq:spatDiscr}. We carry out the final step of computing the time evolution by using a total variation diminishing implementation of the Runge-Kutta algorithm \cite{Shu:1988} of the same order as the spatial reconstruction.
If $\ell_j$ is the spatial operator that represents the right hand side of either of \eqref{eq:spatDiscrB}--\eqref{eq:spatDiscr} and $u_j$ the respective field, the evolution scheme for every cell $j$ is given by
\begin{align}
\begin{split}
    &u^{(0)}_j = u^n_j,\\
    &u^{(1)}_j = u^{(0)}_j + \ell_j(u^{(0)})\delta t,\\
    &u^{(2)}_j = \frac{3}{4} u^{(0)}_j + \frac{1}{4} u^{(1)}_j + \frac{1}{4}\ell_j(u^{(1)})\delta t,\\
    &u^{n+1}_j = \frac{1}{3} u^{(0)}_j + \frac{2}{3} u^{(2)}_j + \frac{2}{3}\ell_j(u^{(2)})\delta t.
\end{split}
\end{align}
The time step $\delta t$ has been chosen at every iteration in order to satisfy the Courant-Friedrich-Lewy condition $\delta t < \delta x / \left|v_{max}\right|$ \cite{Leveque:2002}, where $v_{max}$ is the maximum speed at the boundary of any cell for that time step.

Boundary conditions are imposed at the boundaries of the domain. At $x=0$, since $B = xb$ we assume $B(0) = 0$, while $\varepsilon(0) = 0$ to remain consistent with \eqref{epsilon}. For the outer boundary, we assume there is no infalling matter from beyond the outermost lattice point so $\dot{B}(x_N) = 0$ and $\dot{\varepsilon}(x_N) = 0$.

From the solution for $B, \varepsilon$, there are two quantities of interest that we will calculate: the energy density $\rho$ of the dust field, and the function
\begin{equation}
\Theta\equiv\frac{4(1+\varepsilon)}{x^2} \theta_{+}\theta_{-},
\end{equation}
whose zeros indicate the location of apparent horizons, and where $\theta_+$ and $\theta_-$ denote the expansion of outgoing and ingoing radial null rays respectively. We use the following relations to compute these two quantities:
\begin{equation}\label{eq:rho_comp}
    \rho(x,t) = -\frac{1}{4\pi x^2}\left( \dot{B}+ \frac{x}{2}\partial_x \varepsilon\right),
\end{equation}
\begin{equation}\label{eq:thetaplus}
    \Theta = 1 - \frac{x^2}{4 (1 + \varepsilon)} \sin^2 \frac{2B}{x^2}.
\end{equation}

A final point to discuss is due to the approximation done to the integral \eqref{eq:approxInt}, that leads to a small error that slowly accumulates in the solution for $\varepsilon$ (and also $B$, as $\varepsilon$ is a source in the equation of motion for $B$). In the case of an open or flat interior ($\varepsilon \ge 0$), this error has no impact on the main features of the numerical results for the dynamics that agree extremely well with analytical results---the only impact is a small quantitative error on quantities like the black hole lifetime.

On the other hand, if the interior region has positive spatial curvature ($\varepsilon < 0$) then the error due to the approximation in handling \eqref{eq:approxInt} causes the spatial curvature to be too large, and in the post-bounce phase, when the error grows sufficiently this causes the shock to split, with part of the shock recollapsing and the other part continuing to move outwards. An analysis of the discontinuity in $B$, $\varepsilon$, and the energy density $\rho$ reveals that this split is due to a smoothing of the discontinuity in $\varepsilon$ that leads to the source term in the $B$ equation being smaller than it should be, decreasing the value of $B$. Since $B$ determines the velocity $v_B$ of the field, the decrease in $B$ leads to $v_B$ becoming negative in a portion of the innermost part of the shock, thereby causing the shock to split.

\begin{figure}[t]
    \centering
    \includegraphics[width=\textwidth]{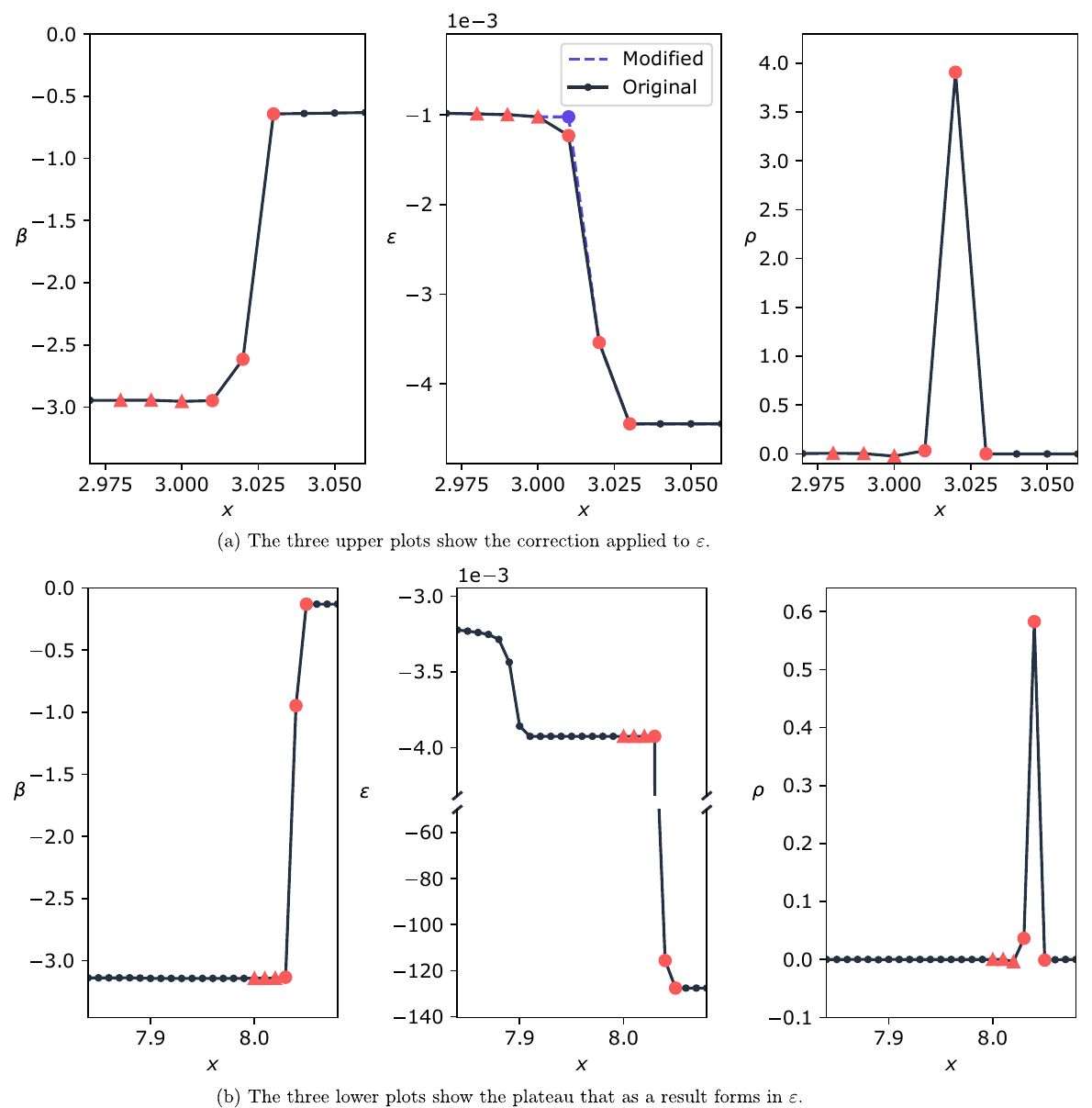}
    \caption{\footnotesize An example of the correction applied to the $\varepsilon$ field is shown in the top three plots. The energy density $\rho$ is used to identify the position of the shock, denoted by the three red circular dots. The value of $\varepsilon$ at the innermost of these three points is corrected to be aligned with its value at nearby points lying within the shock (denoted by red triangles), mimicking that property of $\beta = B/x^2$ just inside the shock. This correction is only applied to simulations with positive spatial curvature $\varepsilon < 0$, and only starting from a short time after the bounce. This modification of $\varepsilon$ is a slight over-correction, so spatial derivatives in $\varepsilon$ near the correction point are smaller than they should be, causing the dynamics to freeze at this point, as shown in the bottom three plots. As the value of $\varepsilon$ at points to the right also approach the same value, they freeze as well, leading to the formation of a plateau, rather than continuing to decrease towards zero. This is a small effect, in this example the value of $\varepsilon$ plateaus near $-4 \times 10^{-3}$ instead of continuing to decrease to values of the order of $-3 \times 10^{-3}$ (as compared to values of $-130 \times 10^{-3}$ outside the shock), and only has a small quantitative effect on the computation of, for example, the black hole lifetime, but does not affect the main qualitative features of the dynamics.}
    \label{fig:combinedFigure}
\end{figure}

To fix this error, for simulations with $\varepsilon < 0$ we perform a small manual correction to $\varepsilon$, as shown in Fig.~\ref{fig:combinedFigure}(a), to properly align the discontinuity in $\varepsilon$ with the discontinuity in $B$. Specifically, the shock lies between three points on the radial lattice, with the innermost point $x_\mathcal{I}$ having very nearly the same value of $\beta = B/x^2$ as at other nearby points $x < x_\mathcal{I}$. The same should be true for $\varepsilon$, but is not due to the error arising from the approximation used to handle \eqref{eq:approxInt}, so to fix that we manually correct the value of $\varepsilon(x_\mathcal{I})$ to be equal to the point immediately to its left; this correction is applied at each time step, but only after the bounce, and fully solves this problem of an incorrect recollapse. (This correction is not applied to simulations with $\varepsilon \ge 0$.)

This procedure, however, gives a slight over-correction to the value of $\varepsilon(x_\mathcal{I})$, and therefore the spatial derivative in $\varepsilon$ at $x=x_\mathcal{I}$ is smaller than it should be. This causes $\varepsilon$ to form a plateau for points lying inside the shock but outside the first point where this correction was made, as shown in Fig.~\ref{fig:combinedFigure}(b), while $\varepsilon$ should be expected to continue to decrease and near zero as the shock moves outwards. Importantly, although this plateau in $\varepsilon$ is a small source of numerical error in the computation of quantitative predictions like the black hole lifetime, it does not affect the qualitative dynamics which, with the correction to $\varepsilon$ described above, agree very well with analytical results as shall be seen below.

This completes the description of the procedure we use to numerically determine the dynamics; the Fortran code we used is available online \cite{github}.

\subsection{Configurations with a sharp boundary}
\label{s.num-sharp}

\begin{figure}
    \centering
    \includegraphics[scale=0.44]{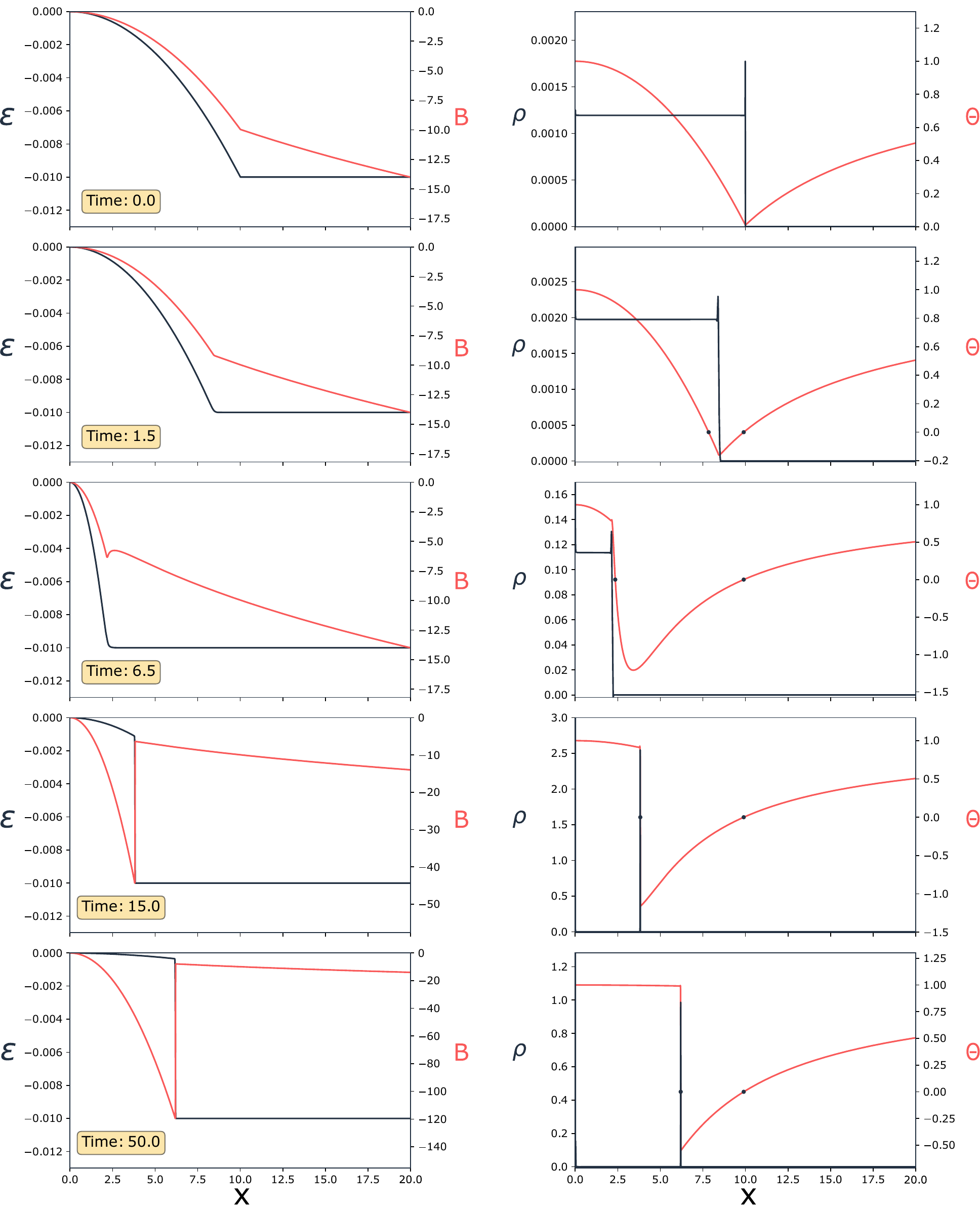}
    \makeatletter\long\def\@ifdim#1#2#3{#2}\makeatother
    \caption{\footnotesize
    Frames from a simulation with an initial configuration with a sharp boundary given by \eqref{rhocont} for $\rho$ and \eqref{epsilon} for $\varepsilon$, with parameters $M=5, \alpha = 0.01, x_0=10$. Each row shows a different time: the top two rows are during the collapse, the third during the bounce and the bottom two during the expansion. The left column shows the $\varepsilon$ field in black and the $B$ field in red, while the right column shows the energy density $\rho$ in black and the function $\Theta$ in red; the black dots show the zeros of $\Theta$ corresponding to the location of the apparent horizons.}
    \label{fig:closedOSframes}
\end{figure}

We start by considering configurations with a sharp boundary, as the numerical results for these configurations can be directly compared with the analytic results obtained in Sec.~\ref{s.sharp}. The initial configuration for $B$ is given by $B = xb$ where $b$ is set by \eqref{eq:IDb}, while $\varepsilon$ is fixed by \eqref{epsilon}. We set $x_1 - x_0 = 2\delta x$, and we take the lattice spacing in the radial direction to be $\delta x = 0.01$. We consider a range of values for $M$ and $\alpha$, where $M = \int_0^{x_1} \dd \tilde x \: \tilde x^2 \rho(\tilde x, t_0)$ is the total gravitational mass. Specifically, we performed runs with different $M$ lying in the interval between $2m_{\text{Pl}}$ and $10m_{\text{Pl}}$ (exploring larger masses is not feasible due to the computational expense associated with solving the dynamics, especially since the black hole lifetime scales as $M^2$), and for both positive and negative values of $\alpha$ ranging from $-0.6$ to $0.6$, while we set $x_0 = 2M$ so (given the small tail between $x_0$ and $x_1$) initially there are no horizons.

Representative frames from a typical simulation for $M=5m_{\text{Pl}}$ and $\alpha = 0.01$ are shown in Fig.~\ref{fig:closedOSframes}. The left column shows the fields $B$ in red and $\varepsilon$ in black at five instants of time, while the right column shows the dust energy density in black and the function $\Theta$ in red at those same instants of time; the roots (if any) of $\Theta$ give the location of the inner and outer apparent horizons and are indicated by black dots.

The dynamics agrees extremely well with the expectations from analytic calculations: the ball of dust collapses and the energy density in the interior region grows until it reaches a critical value, at which point there is a bounce and the shock wave forms.  The shock then slowly moves outwards, with the energy density and spatial curvature in the interior region both rapidly decreasing. (Note that during contraction the density profile exhibits an anomalous peak at the outer edge of the dust ball, this is a numerical artefact due to the presence of large derivatives in $B$ and $\varepsilon$ and is a byproduct of computing $\rho$ using finite differences in~\eqref{eq:rho_comp}. Importantly, this numerical artefact has no impact on the dynamics since $\rho$ does not appear in the equations of motion.) Also, for a given value of $M$, we find that the location of the outer horizon, as found by the numerics, is in excellent agreement with the analytical prediction. Throughout the entire evolution, the two calculations agree up to an error of $\delta x$, and for the simulations we ran this error can be reduced by four orders of magnitude simply by using common root finding techniques on \eqref{eq:thetaplus}.

An important observable for non-singular black holes like the ones studied here is their lifetime $T$, as observed by a distant stationary observer who detects light rays emitted by the star just before its radius becomes smaller than the Schwarzschild radius and it becomes a black hole, and later detects light rays emitted from the shock immediately after the shock exits the horizon. This observed lifetime is equal to the coordinate time $t$ elapsed between the formation and vanishing of the horizons \cite{Kelly:2020lec}. For configurations with a sharp boundary, analytic calculations give the result \eqref{an-life}, predicting that $T$ is proportional to $M^2$ and has a somewhat complicated dependence on $\alpha$.

\begin{figure}
  \centering
  \includegraphics[width=\textwidth]{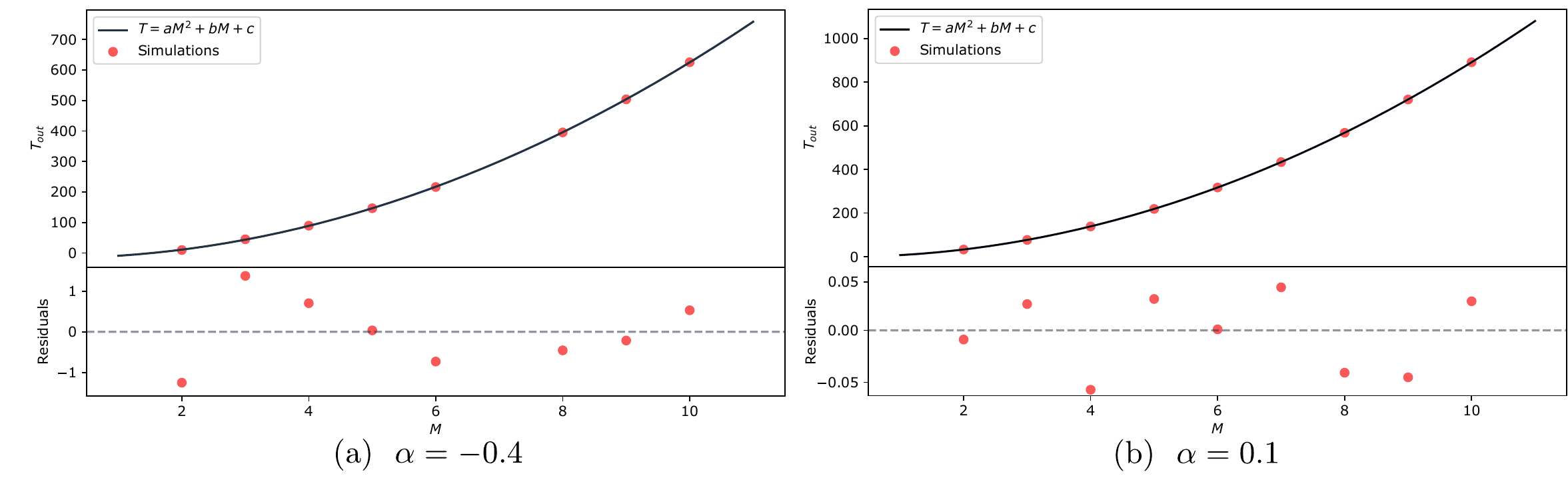}
  \caption[]{\footnotesize
  The lifetime of a black hole as a function of the gravitational mass $M$, for configurations with a sharp boundary. The red dots are the results of numerical simulations for different values of $M$ and fixed $\alpha$, while the black curve is the best quadratic fit. For $\alpha = -0.4$ the best fit is $T = 6.46 M^2 -1.21 M -0.36$, while for $\alpha = 0.1$ the best fit is $T = 9.06M^2 -1.45 M - 0.35$.}
  \label{fig:timeVsM}
\end{figure}

We show the lifetime as a function of $M$ in Fig.~\ref{fig:timeVsM} for two representative values of $\alpha$, an example of positive spatial curvature with $\alpha = 0.1$, and an example of negative spatial curvature with $\alpha = -0.4$. In both cases, we fit the dependence of $T$ on $M$ to a quadratic relation, finding an excellent fit. Notably, the coefficient to $M^2$ in this fit, denoted by $a$ in the plots, agrees up to the second decimal digit with the analytical value derived from \eqref{an-life} for the respective values of $\alpha$: for $\alpha = 0.1$ we obtain $9.06 M^2 + O(M)$, and we get $6.46 M^2 + O(M)$ for $\alpha = -0.4$.

\begin{figure}
  \centering
  \includegraphics[width=\textwidth]{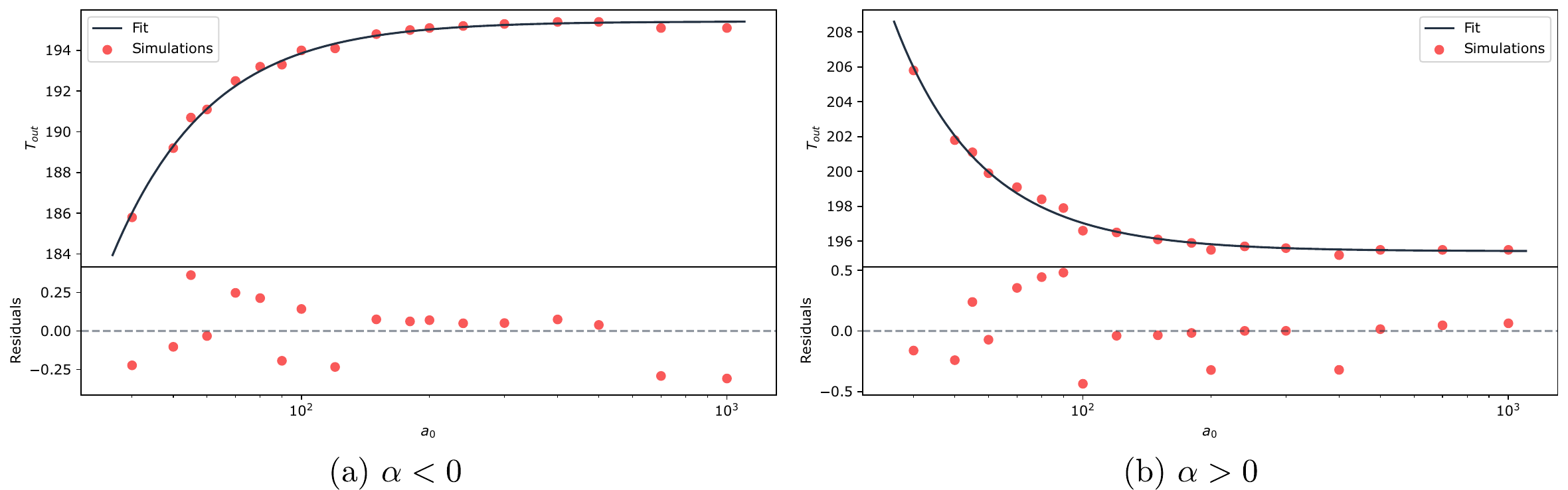}
  \caption[]{\footnotesize
  The lifetime of a black hole as a function of $a_0 = \sqrt{k x_0^2 / \alpha} $, for the case $M=5$. 
 The red dots are the results of numerical simulations, while the black curve is the best fit to the function \eqref{fit-alpha}, with the only free parameter being the overall prefactor, which analytic calculations give as $\pi R_S^2$. For negative $\alpha$, the best fit for the prefactor is 316, while for positive $\alpha$ it is 320.} 
  \label{fig:timeVsa0}
\end{figure}

The dependence of the black hole lifetime on the parameter $\alpha$ has also been investigated. Keeping the gravitational mass fixed to $M=5m_{\text{Pl}}$, the parameter $a_0 = \sqrt{k x_0^2 / \alpha} $ was varied within the range of $40$ to $1000$, with the results shown in Fig.~\ref{fig:timeVsa0}.
These numerical results can be compared to the analytic calculations of Sec.~\ref{s.sharp}, to do this we fit the numerical results to the function
\begin{equation} \label{fit-alpha}
T = {c} \cdot \frac{2\ln\left(1-\alpha\right)+\left(1+\alpha\right)^{2}-1   }{\left(-\alpha \right)^{3}},
\end{equation}
with $c$ as the free parameter, which according to the analytic calculations should be given by $\pi R_S^2$ (which for $M = 5 m_{\rm Pl}$ is $100\pi$). For negative $\alpha$ we find ${c} = 316$, while for positive $\alpha$ we find ${c} = 320$, both of the values agree with the analytic prediction to within an error of $2\%$.
There is a higher accuracy for the case of a negative spatial curvature, which we attribute to the larger numerical error, and the additional corrections needed to the code, for runs with positive spatial curvature, as discussed in Sec.~\ref{s.num}.

The black hole lifetime increases with $\alpha$, as expected, and in the limit of $\alpha$ close to 0, the two plots tend to the same value $T = 195.45$, which also agrees with what is found for the spatially flat $\alpha=0$ case \cite{Husain:2021ojz, Husain:2022gwp}. There are two other limits of interest: $\alpha$ being very negative, or very close to 1. In the first of these two limits, for negative $\alpha$ the lifetime decreases, and eventually vanishes---this is because, at fixed $M$, for sufficiently negative $\alpha$ no horizons ever form, as can be seen in~\eqref{eq:exactHor} and Fig.~\ref{fig:deltaVSxtilde}, and then the lifetime by definition is zero. On the other hand, for $\alpha$ close to 1, the black hole lifetime can become arbitrarily large (keeping $M$ fixed), although it will always remain finite for $\alpha < 1$.

Finally, we also checked the prediction that for $\alpha > 0$ the shock wave asymptotically approaches the maximal value $R_S / \alpha$, see~\eqref{maxL}, rather than recollapsing as would be expected from Oppenheimer-Snyder collapse models that do not include local degrees of freedom (and therefore do not allow for the possibility of a shock to form). Due to the limitations of numerical simulations with a finite runtime, it is not possible to directly check asymptotic predictions like this one. Nonetheless, by running the code for as long as the numerics remains stable (while maintaining a reasonable trade-off between the lattice spacing and the total computational time), we were able to check that during the entire runtime after the bounce, the shock continues to move outwards at an ever-decreasing velocity, and this continues to be true well beyond the time $t_{OS}$ that the Oppenheimer-Snyder model predicts a recollapse---specifically, we were able to verify this for a time after the bounce of $\sim 3.5 \, t_{OS}$.

In summary, the numerics for configurations with a sharp boundary agree very well with the analytic results for the collapse phase, for the bounce, and for the formation of the shock, as well as for the outward movement of the shock, for both positive and negative $\varepsilon$. Nevertheless, a more rigorous approach to handle the term \eqref{eq:approxInt} could provide even more accurate simulations.

\subsection{Results for more general initial profiles}

\begin{figure}
    \centering
    \includegraphics[scale=0.44]{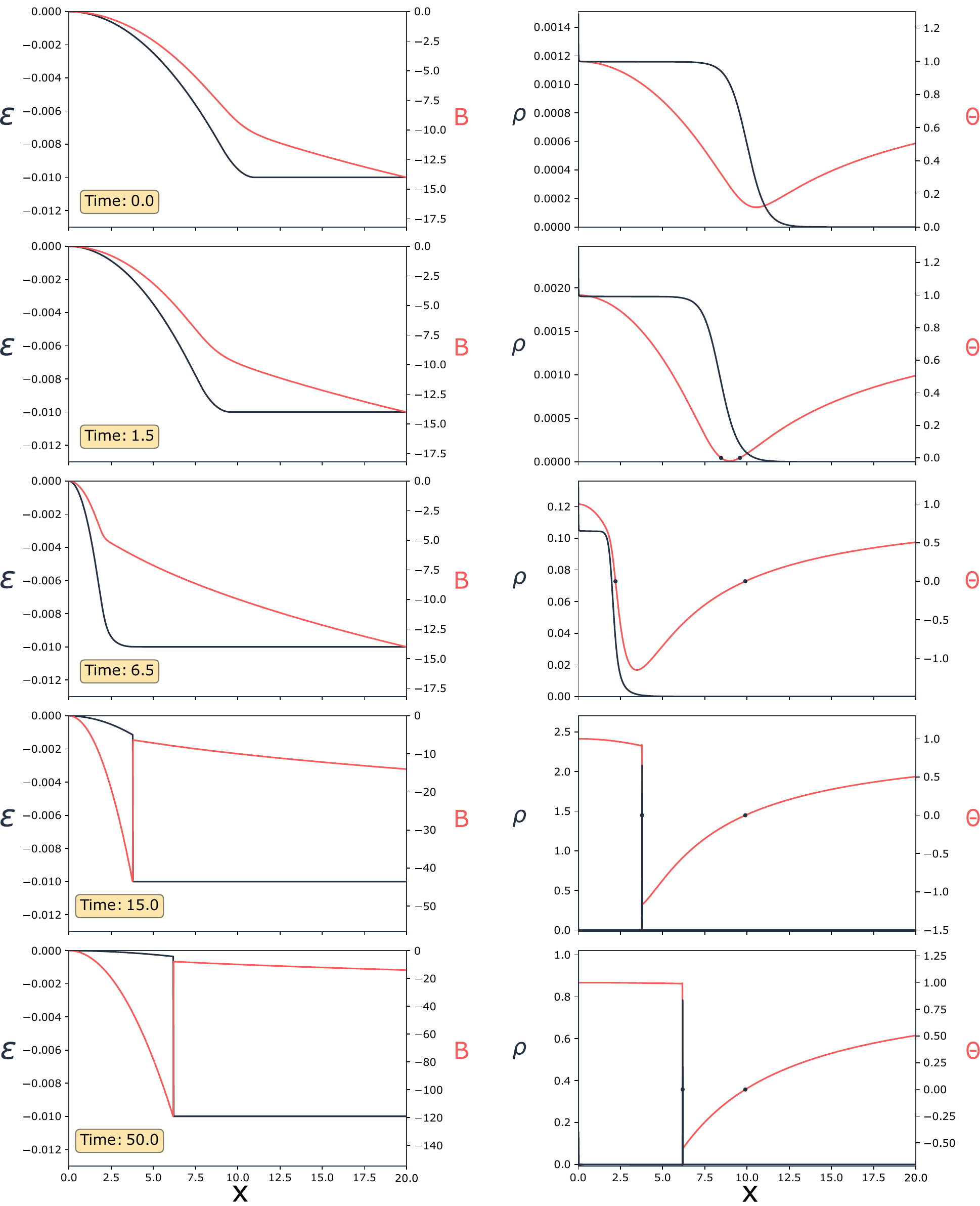}
    \caption[]{\footnotesize
    Frames from a simulation with an initial configuration with a smooth boundary given by \eqref{eq:tanhRho} for $\rho$ and \eqref{epsilon} for $\varepsilon$, with parameters $M=5, \alpha = 0.01, x_0 = 10, \sigma=1.1$. Each row shows a different time: the top two rows are during the collapse, the third during the bounce and the bottom two during the expansion. The left column shows the $\varepsilon$ field in black and the $B$ field in red, while the right column shows the energy density $\rho$ in black and $\Theta$ in red; the black dots show the zeros of $\Theta$ corresponding to the location of the apparent horizons.}
    \label{fig:closedtanhframes}
\end{figure}

We also considered a variety of other types of initial configurations for $\rho$ and $\varepsilon$, which together determine the initial configuration for $B$ through \eqref{eq:IDb} and $B = xb$.

For example, it is possible to consider an initial configuration for the energy density $\rho$ 
\begin{equation}\label{eq:tanhRho}
    \rho(x) = C \left( 1 - \tanh \frac{x-x_0}{\sigma} \right),
\end{equation}
where $C = M / \int_0^\infty \dd x ~ x^2 (1 - \tanh[(x-x_0)/\sigma])$. For $\varepsilon(x,t=0)$ we took \eqref{epsilon} up to the small modification of, between $x_0 \pm 10 \, \delta x$, using a third-degree polynomial to interpolate between the two regimes to ensure that $\varepsilon$ is also continuous and differentiable everywhere (different intervals were considered for the interpolation, the choice had a negligible impact on the resulting dynamics). The results of this run are shown in Fig.~\ref{fig:closedtanhframes} for $M=5, \alpha = 0.01, x_0=10, \sigma = 1.1$.

This initial configuration is very similar to the one studied in Sec.~\ref{s.sharp} and Sec.~\ref{s.num-sharp} (and the results shown in Fig.~\ref{fig:closedtanhframes} can be directly compared to Fig.~\ref{fig:closedOSframes}), except with a smoother decrease in $\rho$ at the boundary of the star. As can be seen by comparing the results, the qualitative dynamics are very similar, with a bounce soon followed by the formation of a shock that slowly moves outwards. There are a few differences which consist of quantitative details; for example, at time $t=1.5$ the minimum of $\Theta$ is slightly smaller for the sharper configuration. Other than small quantitative details like these, the qualitative results for these two types of configurations are very similar.

\begin{figure}
    \centering
    \includegraphics[scale=0.44]{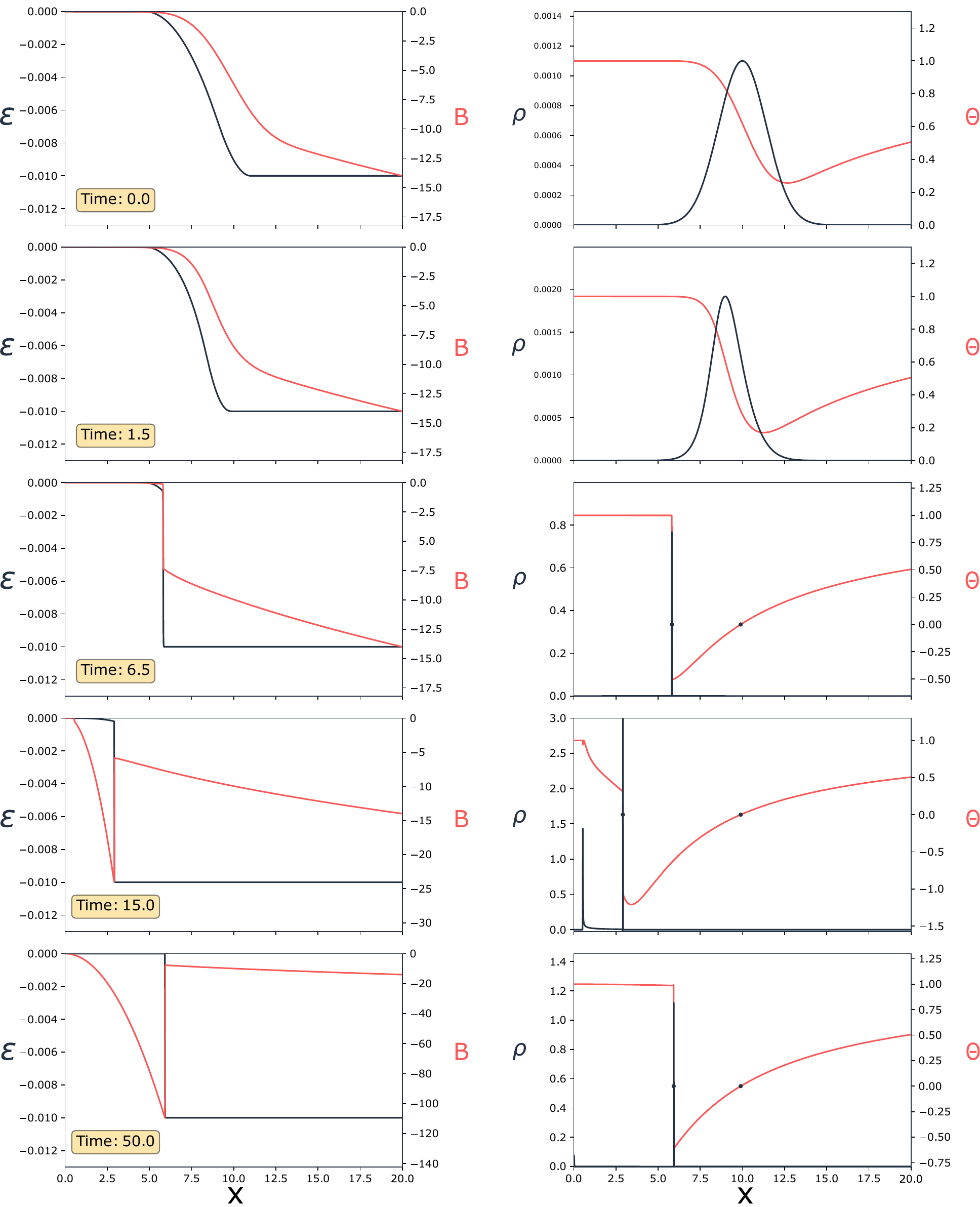}
    \caption[]{\footnotesize
    Frames from a simulation with an initial configuration corresponding to an infalling Gaussian wave packet given by \eqref{eq:expRho} for $\rho$ and \eqref{epsiloooon} for $\varepsilon$, with parameters $M=5, \alpha = 0.01, x_0 = 10, \sigma=2$. Each row shows a different time: the top two rows are during the collapse, the third during the bounce and the bottom two during the expansion. The left column shows the $\varepsilon$ field in black and the $B$ field in red, while the right column shows the energy density $\rho$ in black and $\Theta$ in red; the black dots show the zeros of $\Theta$ corresponding to the location of the apparent horizons.}
    \label{fig:closedSGframes}
\end{figure}

Many more types of initial configurations can be studied. One possibility is a collapsing dust wave packet, whose energy density initially has a Gaussian shape,
\begin{equation} \label{eq:expRho}
    \rho(x) = C \exp{\left[- \frac{(x-x_0)^2}{\sigma^2}\right]},
\end{equation}
where the overall normalization $C$ is proportional to the total mass $M$. Note that since $\rho$ is initially very nearly vanishing close to the origin, in the closed case where $\varepsilon < 0$ it is necessary for the spatial curvature to be sufficiently small so that the argument of the square root in \eqref{eq:IDb} is positive for all $x$. To respect this condition, here we set
 \begin{equation}
     \varepsilon(x, t=0) = \begin{cases}
        0 \quad &\text{if $x<x_\ell$,}\\ \displaystyle
        -\alpha \cdot \frac{(x - x_\ell)^2} {(x_0 - x_\ell)^2} \quad &\text{if $x_\ell<x<x_0$,}\\ \displaystyle
        -\alpha \quad & \text{if $x>x_0$,}
        \label{epsiloooon}
     \end{cases}
\end{equation}
where $\alpha$ is a constant, and $x_\ell$ is chosen to be the smallest possible value that ensures that the argument of the square root in \eqref{eq:IDb} remains positive; $x_\ell$ is found numerically for each run during the computation of the initial data. Again, cubic interpolation has been used around $x_0$ to remove the discontinuity in the derivative of $\varepsilon$.

The results for this set of initial data are shown in Fig.~\ref{fig:closedSGframes} for $C$ chosen so that $M=5$, and $\alpha = 0.01$, $x_0 = 10, \sigma=2$. These are qualitatively similar to the other types of initial conditions we have considered so far: the Gaussian shell collapses inward and the width of the profile narrows during the collapse, then there is a bounce and the formation of a shock wave that slowly moves outwards, eventually emerging from the outer horizon.

\begin{figure}
    \centering
    \includegraphics[scale=0.44]{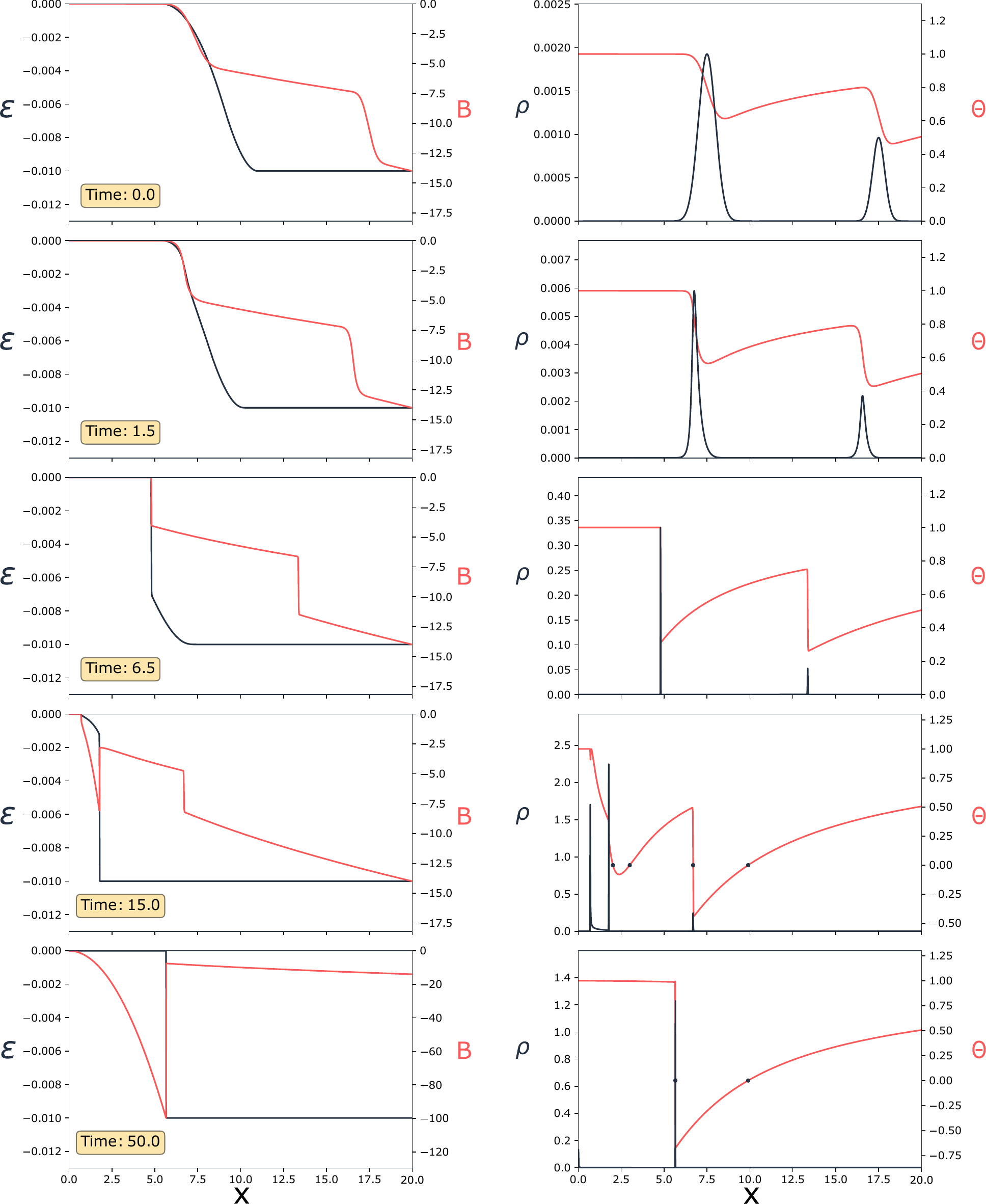}
    \caption[]{\footnotesize
    Frames from a simulation with an initial configuration corresponding to two infalling Gaussian wave packets given by \eqref{eq:dblGauss} for $\rho$ and \eqref{epsiloooon} for $\varepsilon$, with parameters $M=5, \alpha = 0.01, x_0 = 10, x_1 = 7.5, \sigma_1^2 = 0.5, x_2 = 17.5, \sigma_2^2 = 0.25$. Each row shows a different time: the top two rows are during the collapse, the third during the bounce and the bottom two during the expansion. The left column shows the $\varepsilon$ field in black and the $B$ field in red, while the right column shows the energy density $\rho$ in black and $\Theta$ in red; the black dots show the zeros of $\Theta$ corresponding to the location of the apparent horizons.}
    \label{fig:closedDGframes}
\end{figure}

Another initial configuration that is of interest is a double Gaussian packet where
\begin{equation}\label{eq:dblGauss}
    \rho(x) = C \left( 2 \exp{\left[- \frac{(x-x_1)^2}{\sigma_1^2}\right]} + \exp{\left[- \frac{(x-x_2)^2}{\sigma_2^2}\right]} \right),
\end{equation}
and the spatial curvature is given by \eqref{epsiloooon}, where $x_0 = 10$ lies between the two Gaussians; the results obtained for these initial conditions are shown in Fig.~\ref{fig:closedDGframes} for $\alpha = 0.01, x_1 = 7.5, \sigma_1^2 = 0.5,  x_2 = 17.5, \sigma_2^2 = 0.25$ and $C$ is chosen so the total gravitational mass is $M=5$. (It is of course possible to allow different relative factors between the two Gaussian profiles, here for the sake of concreteness we set the relative factor to be 2.) As in the other runs, we observe the same patterns of collapse, bounce and the formation of an outgoing shock wave. Moreover, as already observed in \cite{Husain:2022gwp} for the $\varepsilon=0$ case, the second shell does not cause the outgoing shock to recollapse, and so these dynamics are not affected by the white hole instability problem \cite{Eardley:1974}---rather, the bounce and shock are robust to in-falling dust.

Since the total mass of the system is constant, it follows that another classical instability that is avoided here is mass inflation \cite{Poisson:1989}. This is because the inner horizon, when formed, is not a Cauchy horizon: specifically, in cases where the inner horizon does form, (i) it does not always remain at the same radial location but rather moves inwards (during collapse) and then outwards (after the bounce) with the shock, and (ii) it ceases to exist within a finite span of time, once the shock reaches the outer horizon.

We end with a comment concerning configurations with two (or more) wave packets in $\rho$. For such a configuration, it is possible for the energy density to vanish between the wave packets (or become extremely small). Nonetheless, for such a configuration it is not necessary to impose that $\varepsilon$ be constant in the vacuum (or near vacuum) region between the wave packets (as is required in the asymptotic region where $\rho=0$, as discussed in Sec.~\ref{subsub:vacuum}). The reason for this is that the condition $\partial_x \varepsilon=0$ is required by $\dot \varepsilon = 0$, but in such an intermediate region located between two wave packets it is not necessary to impose $\dot \varepsilon = 0$, and as a result there is no requirement for $\varepsilon$ be constant in that region either.

\section{Summary and conclusion}
\label{s.con}

We have studied the role of spatial curvature in the gravitational collapse of dust in the context of effective LQG, generalizing earlier results that focused on the spatially flat case \cite{Husain:2021ojz, Husain:2022gwp}. Although the quantitative results are affected by the presence of spatial curvature, the qualitative features of the dynamics are very similar: the gravitational collapse continues until the spacetime curvature reaches the Planck scale, at which point there is a bounce with the dust starting to move outwards, and a shock wave forms (at the latest $\sim t_{\rm Pl}$ after the bounce), with a discontinuity in the gravitational fields as well as the dust energy density. This occurs for a variety of initial configurations, and for positive and negative spatial curvature.

Interestingly, keeping the spatial curvature fixed, in the case of an initial configuration corresponding to a homogeneous interior with a sharp boundary, the lifetime $T$ of a black hole has been found to scale with $M^2$, in agreement with what was found for the spatially flat case \cite{Husain:2021ojz, Husain:2022gwp}, although there is also a dependence on the spatial curvature that can become important for large spatial curvature---see~\eqref{an-life} for details. In fact, numerics show that this dependence holds for a wide range of configurations, where the lifetime depends on the mass $M$ and the (constant) value of the spatial curvature in the exterior vacuum region. After the shock exits the horizon, the shock continues moving outwards, although at a continually decreasing velocity. The shock will reach arbitrarily large radii for the cases of negative and vanishing spatial curvature, while it will asymptotically approach a maximal radius if the spatial curvature is positive (in contrast to expectations based on studies of the Oppenheimer-Snyder model without a shock that predict a cyclic `pulsating star' \cite{BenAchour:2020mgu, Cafaro:2024}).

There are also two limiting cases that are of particular interest. First, it is possible to recover a spatially curved FLRW cosmology by taking the homogeneous limit, and the result agrees exactly with the results of loop quantum cosmology (specifically the `K' loop quantization). This result demonstrates the robustness of the results of LQC, and shows that the same physics is being captured here in the more general context of spherical symmetry. Second, the vacuum exterior region is also of interest, with it closely resembling what was observed in the spatially flat case \cite{Kelly:2020uwj}. In particular, there are two Killing horizons (except for sufficiently small masses of the order of the Planck mass) and the corrections to the Schwarzschild metric are suppressed by a factor of $\Delta/x^2$. Nonetheless, despite their similarities, it is important to emphasize that vacuum solutions with different spatial curvature are not diffeomorphic to each other (as is the case in classical general relativity); rather, these are different vacuum solutions with the same mass, with quantum violations of the no-hair theorem of general relativity.

Finally, since quantum gravity effects have a major impact on the ultimate fate of gravitational collapse---replacing the crushing singularity that would form in general relativity with an outgoing shock wave---it is natural to expect that this will have important ramifications for the information loss problem, and on black hole thermodynamics more generally. In particular, the black hole lifetime that is predicted to be proportional to $M^2$ suggests that the black hole will expire at a time when Hawking evaporation remains entirely subdominant (with a total evaporated mass of the order of $m_{\rm Pl}$), so long as $\alpha$ is not fine-tuned to a value extremely close to 1. Further, since the horizons are not eternal, it may be possible to purify the small amount of Hawking radiation that has occurred. We leave a detailed exploration of these questions for future work.

\appendix

\section{Formation of shell-crossing singularities}
\label{appendix:SCSing}

Shell-crossing singularities can form in LTB spacetimes, due to some dust shells overtaking others; for an analysis in the context of general relativity see \cite{Hellaby:1985}. This is studied most easily using the comoving coordinate $R$, which moves with the dust (and keeping the same time and angular coordinates used for generalized Painlev\'e-Gullstrand coordinates). The relation between these two sets of coordinates is captured by the function $x(R, t)$ that gives the value of the areal radius $x$ for the dust shell located at the comoving radius $R$, as a function of time.

Using comoving coordinates, the dynamics greatly simplifies as $m(R,t)$ and $\varepsilon(R,t)$ are now constants of the motion in terms of these coordinates,
\begin{equation} \label{constmass}
\partial_{t}m(R,t)=0, \qquad
\partial_{t}\varepsilon(t,R)=0,
\end{equation}
and the dynamics are entirely captured by the equation of motion for $x(R,t)$
\begin{equation}
\left( \frac{\Dot{x}}{x}  \right)^{2}=\left(\frac{2Gm}{x}+\frac{\varepsilon}{x^{2}} \right) \left[ 1-\Delta \left(\frac{2Gm}{x}+\frac{\varepsilon}{x^{2}} \right)  \right],
\end{equation}
an ordinary differential equation to be solved at each radius $R$. (Note however that despite the simplification obtained in going from partial differentials equations to ordinary differential equations, solving these equations for $\varepsilon \neq 0$ analytically remains challenging---on the other hand, numerical solutions are more readily attainable.) These dynamics can be derived directly from a loop quantization based on the comoving coordinate $R$ \cite{Giesel:2023tsj, Giesel:2023hys}, or by a coordinate transformation from the effective dynamics expressed in terms of the generalized Painlev\'e-Gullstrand coordinates \cite{Fazzini:2023ova}.

It is important to emphasize that although the use of the comoving coordinate $R$ gives dynamics for each individual dust shell that is decoupled from its neighbours, this coordinate choice only remains valid so long as the dust shells do not cross: if the shells do cross, then the comoving coordinate $R$ fails and it is necessary to use a different set of coordinates. This is a common feature of non-linear wave equations, and a rich mathematical framework has been been developed to handle such a shell-crossing (known in the mathematical literature as a characteristic crossing) that signals the formation of a discontinuity in the field and therefore the need to find weak solutions (i.e., solutions that are not necessarily everywhere continuous or differentiable) to the dynamics.

In LTB spacetimes, shell-crossing singularities are of particular interest; these are shell-crossings where the curvature scalars diverge due to the energy density of the dust field diverging when dust shells cross. The dust energy density is given by \cite{Fazzini:2023ova}
\begin{equation} \label{def-rho}
 \rho(R,t)=\frac{\partial_{R}m}{4 \pi x^{2} \partial_{R}x },
\end{equation}
(this is the same expression as in general relativity \cite{Hellaby:1985}) and it is clear that shell-crossing singularities occur at any $R$ for which $\partial_R m \neq 0$ and at some time $\partial_R x$ vanishes. Note that since $m(R,t)$ is independent of time, the numerator in $\rho(R,t)$ is also independent of time. As a result, the question of whether $\rho$ diverges at some radius $R$ (for which initially $\partial_R m \neq 0$) reduces to calculating whether $\partial_R x$ ever vanishes.

In the $\varepsilon=0$ marginally bound case, for many initial configurations it is possible to solve the dynamics and calculate $\rho$ analytically \cite{Fazzini:2023ova}. For $\varepsilon \neq 0$, numerics are needed, and in this case it is sufficient to solve for $x(R,t)$ in the region where $\partial_R m \neq 0$: then, if any curves $x(R,t)$ for $R=R_1$ and $R=R_2$ cross at $t=t_i$, it necessarily follows that $\partial_R x = 0$ for some $R_1 \le R \le R_2$ and some $t \le t_i$, at which spacetime point there is a shell-crossing singularity.

Physically, the presence of a shell-crossing singularity shows that distinct dust shells lie at precisely the same location, which is what causes the energy density (and therefore curvature scalars) to diverge. It is clear that this is not a coordinate artefact, but rather a truly physical process which signals the onset of a discontinuity in the gravitational field and the need to allow for a weak solution to the dynamics beyond a shell-crossing singularity.

\begin{figure}
    \centering
    \includegraphics[scale=0.68]{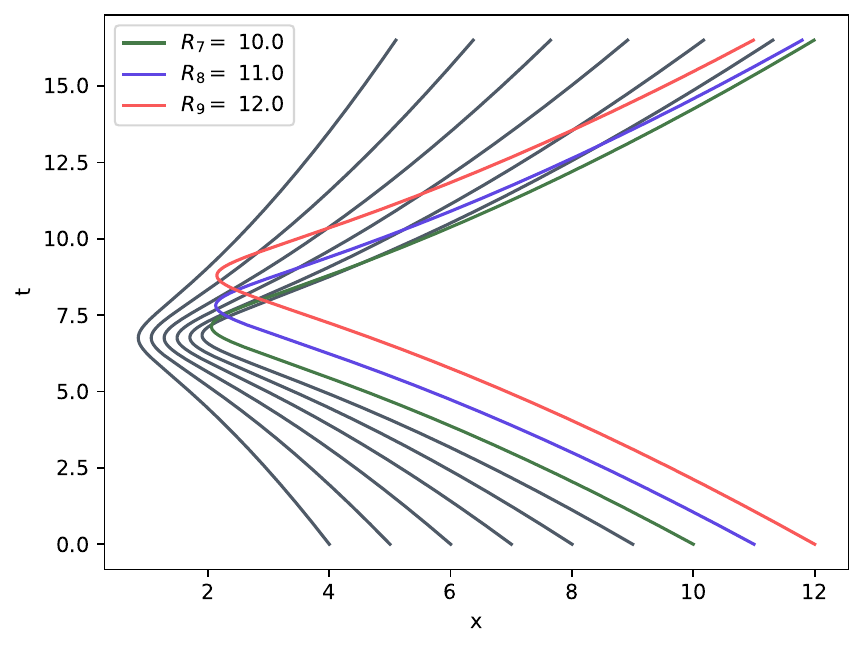}
    \caption{\footnotesize
    Dynamics of many shells. The initial data is that of equations \eqref{eq:rhospec} with $R_0 = 10$, $\sigma=1.1$, $\alpha = 0.01$ and $C$ proportional to the total mass $M=5$. All shells have strictly positive initial energy density. The black curves describe the evolution of shells on the plateau of the distribution of the initial energy density, while the others depict shells on the tail. As expected, the black shells do not intersect, whereas the green, blue and red shells cross the black ones immediately after the bounce of the latter, in agreement with the production of the shockwave in the integral equations and with expectations from the marginally bound case where shell-crossing singularities typically occur in regions where the dust energy density varies sufficiently rapidly \cite{Fazzini:2023ova}. Note that a single shell crossing (between shells where $\partial_R m \neq 0$) is sufficient to demonstrate the presence of a shell-crossing singularity.}
    \label{fig:shellDyn}
\end{figure}

We considered each of the configurations that were studied in this paper to determine whether a shell-crossing singularity occurs, with the result that a shell-crossing singularity was found to occur in every single one of these configurations. In particular, this includes the configurations with a homogeneous interior and a sharp boundary studied in Sec.~\ref{s.sharp}.

As a representative example, here we present the results for the initial configuration given by
\begin{equation} \label{eq:rhospec}
   \rho(R,t_{0}) = C \left( 1 - \tanh \frac{R-R_0}{\sigma} \right), \qquad
   \varepsilon(R,t_{0})= \begin{cases}
    \displaystyle -\alpha  \cdot\frac{R^{2}}{R_{0}^{2}}, \qquad & \text{for $R<R_{0}$, }    \\
    \displaystyle -\alpha, \quad & \text{for $ R > R_{0}$,}
 \end{cases}
\end{equation}
where $\alpha = k R_0^2 / a(t_0)^2$; these are the same initial conditions as those given in \eqref{epsilon} and \eqref{eq:tanhRho}, although expressed in terms of the comoving radius $R$ instead of the areal radius $x$ (with the initial condition $x(R, t_0) = R$). This is a configuration with a nearly homogeneous interior, whose boundary is approximately located at $R_0$ and has width $\sim \sigma$, and with a long tail outside where $\rho$ asymptotes to zero.

It is easy to calculate $m(R,t_0)$ from $\rho(R, t_0)$ from \eqref{def-rho} by using the initial condition $x(R, t_0) = R$ which implies that $\partial_R x = 1$ at $t=t_i$. Then, $m(R, t) = m(R, t_0)$ follows directly from \eqref{constmass}, and it is clear that for this configuration $\partial_R m \neq 0$ for all $R$. As a result, if any of the curves $x(R, t)$ intersect, then it necessarily follows that there is a shell-crossing singularity (in which case a discontinuity forms in the gravitational field, and it is necessary to look for a weak solution to the dynamics).

Several representative curves $x(R,t)$ for different $R$ are shown in Fig.~\ref{fig:shellDyn}. Shell crossings occur soon after the bounce and since, as explained above, $\partial_R m \neq 0$ everywhere, this implies the presence of at least one shell-crossing singularity. (Note that not all shells cross; a single intersection is sufficient to demonstrate the presence of a shell-crossing singularity.) Therefore, a discontinuity in the gravitational field is formed and it is necessary to consider weak solutions to the dynamics, as has been done in Secs.~\ref{s.sharp} and \ref{numerical_section}.

Although here we have presented the numerical results demonstrating the formation of a shell-crossing singularity for the specific set of initial conditions \eqref{eq:rhospec}, using exactly the same numerical test we have also examined each of the families of initial data we considered in this paper, and found that a shell-crossing singularity forms (at the latest a short time $\sim t_{\rm Pl}$ after the bounce) for all of these sets of initial data, in agreement with the observation of the formation of a shock in Secs.~\ref{s.sharp} and \ref{numerical_section}.

\acknowledgments

The work of F.F.~and E.W.-E.~is supported in part by the Natural Sciences and Engineering Research Council of Canada.

\raggedright

\end{document}